\documentclass[%
reprint,
superscriptaddress,
 amsmath,amssymb,
prb,
onecolumn,
]{revtex4-2}
\pdfoutput=1

\usepackage{graphicx}% Include figure files
\usepackage{hyperref}% add hypertext capabilities
\begin{document}

%\preprint{}

\title{Devolatilization of Subducting Slabs, Part II: \\
Volatile Fluxes and Storage}


\author{Meng Tian}
\email[]{meng.tian@csh.unibe.ch}
%\thanks{}
\altaffiliation[Now at: ]{Center for Space and Habitability, University of Bern, Bern, Switzerland}
\affiliation{Department of Earth Sciences, University of Oxford, South Parks Road, Oxford, OX1 3AN, UK.}

\author{Richard F. Katz}
%\email[]{}
%\thanks{}
\affiliation{Department of Earth Sciences, University of Oxford, South Parks Road, Oxford, OX1 3AN, UK.}

\author{David W. Rees Jones}
%\email[]{}
%\thanks{}
\affiliation{Department of Earth Sciences, University of Oxford, South Parks Road, Oxford, OX1 3AN, UK.}
\affiliation{Department of Earth Sciences, Bullard Laboratories, University of Cambridge, Madingley Road, Cambridge, CB3 0EZ, UK.}
\affiliation{School of Mathematics and Statistics, University of St Andrews, North Haugh, St Andrews, KY16 9SS, UK}

\author{Dave A. May}
%\email[]{}
%\thanks{}
\affiliation{Department of Earth Sciences, University of Oxford, South Parks Road, Oxford, OX1 3AN, UK.}


%\date{\today}

\begin{abstract}
Subduction is a crucial part of the long-term water and carbon cycling between Earth's exosphere and interior. However, there is broad disagreement over how much water and carbon is liberated from subducting slabs to the mantle wedge and transported to island-arc volcanoes. In the companion paper Part I, we parameterize the metamorphic reactions involving H$_2$O and CO$_2$ for representative subducting lithologies. On this basis, a two-dimensional reactive transport model is constructed in this Part II. We assess the various controlling factors of CO$_2$ and H$_2$O release from subducting slabs. Model results show that up-slab fluid flow directions produce a flux peak of CO$_2$ and H$_2$O at subarc depths. Moreover, infiltration of H$_2$O-rich fluids sourced from hydrated slab mantle enhances decarbonation or carbonation at lithological interfaces, increases slab surface fluxes, and redistributes CO$_2$ from basalt and gabbro layers to the overlying sedimentary layer. As a result, removal of the cap sediments (by diapirism or off-scraping) leads to elevated slab surface CO$_2$ and H$_2$O fluxes. The modelled subduction efficiency (the percentage of initially subducted volatiles retained until $\sim$200 km deep) of H$_2$O and CO$_2$ is increased by open-system effects due to fractionation within the interior of lithological layers.
\end{abstract}

%\keywords{}

\maketitle

\section{Introduction}
Earth distinguishes itself from other solar-system planets through its habitability that is maintained by its surface climate. Over geological time, water and carbon modulate the climate through geochemical cycles between Earth's exosphere and interior \citep{Dasgupta:2013aa}. Subduction is a tectonic process that brings altered, near-surface rock into the deep Earth and therefore participates in the long-term geochemical cycles. However, the flux of carbon accompanying subduction into deep Earth is still actively debated. \citet{Dasgupta:2010aa} and \citet{hirschmann18} argue that subducting slabs don't experience significant degassing or partial melting and so sequester their carbon into the deep Earth over geological history. On the other hand, \citet{Kelemen:2015aa} contend that most carbon is liberated from slabs and migrates into the subduction-zone mantle lithosphere, rather than being recycled into the deep mantle.

Detailed field and modelling studies also give disparate views on the fate of subducting carbon. \citet{Kerrick:1998aa, Kerrick:2001aa, Kerrick:2001ab} used thermodynamics to construct petrological phase diagrams for representative lithologies in subduction zones (i.e., hydrothermally altered slab mantle, metabasalts, metasediments). They concluded that all representative lithologies bring a significant amount of $\mathrm{CO_2}$ into the deep mantle, except that clay-rich slab sediments undergo complete decarbonation at forearc depths along hot subduction geotherms. Subsequently, \citet{Gorman:2006aa} evaluated the open-system effects induced by aqueous fluid infiltration on the thermodynamic modelling of subduction-zone dehydration and decarbonation. They concluded that $\mathrm{CO_2}$ liberation is still limited, in spite of $\mathrm{H_2O}$-rich fluid infiltration. Field studies on subducted carbonates, however, suggest considerable carbon release by carbonate dissolution \citep{Frezzotti:2011aa, Ague:2014aa}; \citet{Piccoli:2016aa} show that the dissolved carbon can be re-precipitated within the slab or proximal mantle wedge. Further thermodynamic models considering aqueous ionic species also suggest that carbon release from subducting slabs is significant because dissolution of carbon in the form of organic ions can enhance carbon removal from rocks \citep{Sverjensky:2014aa}. In particular, \citet{Connolly18} show that consideration of non-molecular species can roughly double the carbon solubility in fluids co-existing with sediments subducted along cool geotherms.

Purely thermodynamic models of the fate of subducted carbon are zero-dimensional in that the system is assumed to be closed, with no directional mass transfer. In contrast, the one-dimensional, open-system model by \citet{Gorman:2006aa} treated $\mathrm{H_2O}$ \& $\mathrm{CO_2}$ allowing for vertical fluid migration. Two-dimensional geodynamic models of porous fluid migration indicates substantial fluid migration nearly parallel to subducting slabs \citep{Wilson:2014aa}. This focused flow is caused by the formation of a high-permeability channel in the dewatering layer and a compaction pressure gradient that helps contain liquids in the slab. Given that flow directions within the slab are uncertain and reactive flow is path-dependent, it is important to assess the effect of fluid flow direction on the fluxes of $\mathrm{H_2O}$ \& $\mathrm{CO_2}$ out of subducting slabs. 

In the computational treatment of open-system behaviors, an extra challenge exists when incorporating metamorphic reactions involving volatiles into fluid flow modelling. Since fluid movement constantly changes the bulk composition of each subdomain within the modelled slab, the computational cost may be prohibitively high if a traditional phase diagram calculation is applied to each subdomain repeatedly throughout the model evolution. In the companion paper Part I, we have parameterized the coupled dehydration and decarbonation processes for representative subducting lithologies (i.e., sediments, MORB, gabbro, and peridotite). This light-weight thermodynamic module focuses on the behaviors of $\mathrm{H_2O}$ \& $\mathrm{CO_2}$ and can readily capture the fractionation and infiltration effects typical in open systems. Thus it forms the basis of the efficient reactive flow model in the current study. We note here that the liquid volatile phase in the model is a molecular fluid residing in the H$_2$O--CO$_2$ binary, so it excludes other carbon species from consideration. The limitation of this assumption is discussed in section \ref{sec: discussion} and in the companion Part I.

In this paper, we provide a two dimensional model of reactive fluid flow in subducting slabs. Since the dynamics of within-slab flow remains highly uncertain \citep[e.g.,][]{Faccenda:2009aa, Wilson:2014aa, Plumper:2017aa, Morishige:2018aa}, we prescribe the flow direction in our model and investigate model behavior as a function of this parameter, rather than solving equations for momentum conservation. The model incorporates open-system equilibrium thermodynamics and enables us to assess the  factors controlling slab dehydration and decarbonation. In particular, we find that nearly up-slab fluid flow produces a peak in the volatile flux at subarc depths and might thus be relevant for arc magmatism. We find that a sedimentary layer can act as a cap that absorbs $\mathrm{CO_2}$ released from underlying slab lithologies, if the sediments are not removed during subduction. Furthermore, slab lithospheric mantle, if extensively serpentinized, can cause significant $\mathrm{H_2O}$ \& $\mathrm{CO_2}$ fluxes at subarc depths in warm subduction settings. In all the cases we explored, 20--90~wt\% H$_2$O and 80--100~wt\% CO$_2$ in the slabs subduct beyond a mantle depth of $\sim$200~km.

In the following, we start with a description of our model setup in section~\ref{sec: setup}, followed by the numerics of model solution in section~\ref{method}. Model results with closed-system behavior are presented in section~\ref{sec: closedmodel}, which is included to facilitate comparison of our model with previous work assuming constant bulk compositions. We present results on a reference, open-system model in section~\ref{sec: refmodel}, both for comparison with the closed-system model and to serve as a reference case to investigate parameter sensitivity. The subsequent models presented explore the variability of H$_2$O and CO$_2$ fluxes on slab surface in response to changes in fluid flow direction (section~\ref{sec: flowdirection}), slab age (section~\ref{sec: age}), extent of slab mantle serpentinization (section~\ref{sec: hydration}), and removal of slab surface sediments (section~\ref{sec: diapirism}). For all the open-system models we run, the efficiency of H$_2$O and CO$_2$ subduction into the deep mantle is summarized in section~\ref{sec: efficiency}, which is followed by a discussion (section~\ref{sec: discussion}) of the model limitations.

\section{Model Setup} \label{sec: setup}
\begin{figure}[h!]
\centering
\includegraphics[scale=0.8]{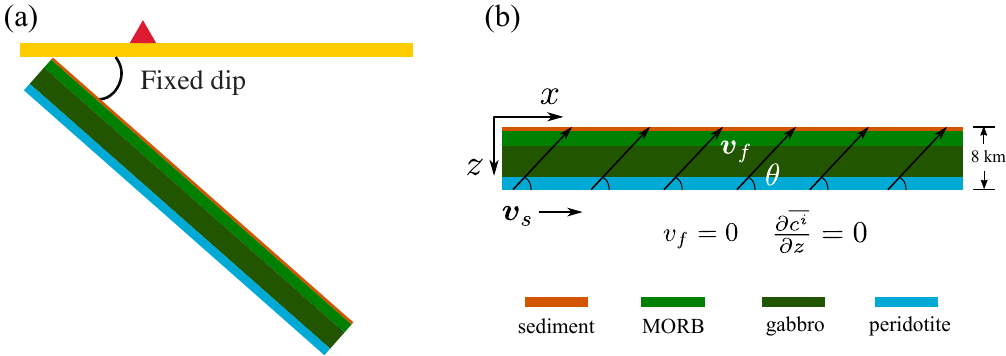}
\caption{Sketch showing the model geometry and boundary conditions. (a) The geodynamic setting of our model. It shows that the model assumes a fixed slab dipping angle and the slab acts like a rigid plate that doesn't deform. A red triangle denotes the position of arc volcano. (b) The modelled slab domain. The $x-$ and $z-$ axes are respectively parallel and normal to the slab extension; this coordinate system is used in all the succeeding figures displaying the entire slab. The slab lithologically consists of four representative rock types as detailed in the legend. Solid velocities within the slab are uniformly set to slab convergence rate ($\boldsymbol{v}_s$). $\theta$ is the angle between $x-$axis and the uniform fluid flow direction. Notations of symbols are listed in Table \ref{tab: symbol} and details on the initial and boundary conditions are provided in section \ref{method}. }
\label{fig: geometry}
\end{figure}

The geometry of the model is illustrated in Figure~\ref{fig: geometry}; we focus only on the subducting slab. Cold temperatures in the slab create strong resistance to viscous (de-)compaction \citep{Wilson:2014aa} and hence the slab experiences negligible isotropic deformation over the model times in this study ($\sim$6 Ma). We therefore treat the slab as a rigid plate in our model. Slab deformation has also been modelled using visco-elastic \citep{Morishige:2018aa} and visco-elasto-plastic rheologies \citep{Faccenda:2009aa}, yielding diverse fluid flow patterns within slab. By treating the slab as a rigid plate, the flow direction can be considered as a free parameter that is to be explored (see below), and we can thus circumvent the uncertainties of slab deformation and coupled fluid flow that are difficult to constrain. 

\begin{figure}[h!]
\centering
\includegraphics[width=0.5\columnwidth, keepaspectratio]{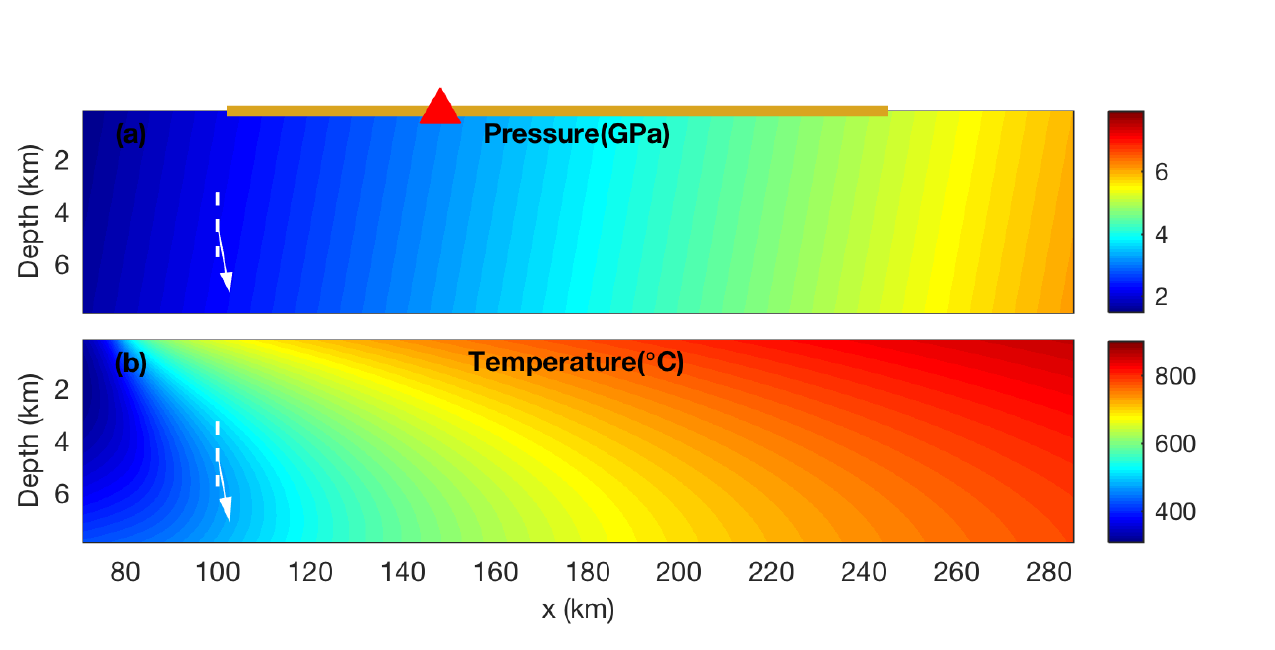}
\caption[Optional caption for list of sub figures]{A representative illustration of the $P$ \& $T$ structure for a 10-Ma-old slab. A convergence rate of 5 cm yr$^{-1}$ and a slab dip of 45$^{\circ}$ are used. The $x$-axis starts from $\sim$70 km because we assume the overriding plate (Fig. \ref{fig: geometry}) has a 50-km-thick lithosphere which corresponds to $\sim$70 km for the starting position of the slab immediately below it. The solid brown line in (a) draws the global range of arc positions projected onto slab surface \citep{Syracuse:2006aa}, and the red triangle marks the site of the average position ($\sim$150 km); the same convention is used in the  following figures. Note that the vertical and horizontal scales are different due to the high aspect ratio of the modelled slab. White dashed lines are a reference direction perpendicular to slab extension and solid white arrows denote the direction of gravity in the stretched domain; the same line and arrow conventions apply in the succeeding figures.}
\label{fig: ptstruct}
\end{figure}

Under the rigid plate assumption, the solid velocities ($\boldsymbol{v}_s$) in the slab are uniformly of magnitude equal to the subduction rate and can be prescribed as a model parameter. If a slab age and dip are further chosen, the steady-state temperature ($T$) and pressure ($P$) of the slab can be calculated using canonical thermo-mechanical models \citep[e.g.,][]{Keken:2008aa}. Our model assumes such a steady-state $P$--$T$ structure of the slab and extracts it from the geodynamic model by \citet{England:2010aa}. Figure~\ref{fig: ptstruct} illustrates an example of this for a 10-Ma-old slab with a convergence rate of 5~\mbox{cm yr$^{-1}$} and a dip of 45$^{\circ}$. This $P$--$T$ field is employed in the following sections except section~\ref{sec: age} where different thermal structures, dependent on slab age, are tested. Note that this is a young, warm, and relatively fast-converging oceanic lithosphere. The relatively fast subduction speed implies a fast spreading rate at mid-ocean ridges, which correlates with small extent of upper mantle serpentinization \citep{Fruh:2004aa}. The chosen H$_2$O content of 1 wt\% for the modelled slab mantle is thus consistent with this spreading rate.

Superimposed on this rigid and thermally steady-state slab is the reactive H$_2$O \& CO$_2$ transport model. This transport model assumes local equilibrium between the solid and percolating liquid phases. The solid phase in the slab is lithologically layered, as shown in Fig.~\ref{fig: geometry}b, and consists of sediments, mid-ocean ridge basalts (MORB), gabbros, and peridotites from top to bottom. In the companion paper Part I, we parameterize the equilibrium partitioning of H$_2$O \& CO$_2$ between the liquid and solid for each of the four representative lithologies; this parameterization serves as a thermodynamic module coupled to the fluid flow model detailed below. 

Following previous studies \citep{McKenzie:1984aa, Scott:1986aa, Katz:2008aa, Keller:2016aa}, mass conservation for liquid and solid phases in porous media are expressed as:
\begin{equation} \label{eq: orgmassl}
\frac{\partial \phi}{\partial t} + \nabla \cdot ( \boldsymbol{v}_f\phi )= \frac{\Gamma}{\rho_f}, 
\end{equation}
\begin{equation} \label{eq: orgmasss}
\frac{\partial (1-\phi)}{\partial t} + \nabla \cdot [ \boldsymbol{v}_s(1-\phi) ]= -\frac{\Gamma}{\rho_s}, 
\end{equation}
where densities of both phases are assumed to be constant, and the meaning of symbols is listed in Table~\ref{tab: symbol}. In addition, the conservation of volatile species $\mathrm{H_2O}$ \& $\mathrm{CO_2}$ in both phases are:
\begin{equation} \label{eq: orgmasslspc}
\frac{\partial (\rho_f\phi c^i_f)}{\partial t} + \nabla \cdot (\rho_f \boldsymbol{v}_f\phi c^i_f) = {\Gamma}_i,
\end{equation}
\begin{equation} \label{eq: orgmasssspc}
\frac{\partial [\rho_s(1 - \phi) c^i_s]}{\partial t} + \nabla \cdot [\rho_s \boldsymbol{v}_s(1 - \phi) c^i_s] = -{\Gamma}_i,
\end{equation}
where $i$ represents either $\mathrm{H_2O}$ or $\mathrm{CO_2}$. Chemical diffusion is neglected here because the P{\'e}clet number ($v_f H/D$) is $\sim$250 if slab thickness ($H$) of 8~km, diffusion coefficient ($D$) of $\sim$$10^{-8}$ $\mathrm{m^2 \, s^{-1}}$, porosity ($\phi$) of $\sim$$10^{-2}$ (e.g., Fig.~\ref{fig: refmodel}g), and a very conservative flux ($v_f \phi$) estimate of $\sim$0.1~m~kyr$^{-1}$ (e.g., Fig.~\ref{fig: refmodel}h) are used. Adding equation~\eqref{eq: orgmassl} to \eqref{eq: orgmasss} and \eqref{eq: orgmasslspc} to \eqref{eq: orgmasssspc} leads to: 
\begin{equation} \label{eq: bulkmass}
-(\rho_s - \rho_f)\frac{\partial \phi}{\partial t} + \rho_f \nabla \cdot ( \boldsymbol{v}_f\phi )= \rho_s \boldsymbol{v}_s \cdot \nabla \phi,
\end{equation}
\begin{equation}\label{eq: bulkspc}
\frac{\partial [\rho_f \phi c^i_f + \rho_s(1 - \phi) c^i_s]}{\partial t} + \rho_f \nabla \cdot [\phi \boldsymbol{v}_f c^i_f] + \rho_s \boldsymbol{v}_s \cdot \nabla [(1 - \phi) c^i_s]= 0.
\end{equation}
Note that, in deriving equations~\eqref{eq: bulkmass} and \eqref{eq: bulkspc}, the zero-compaction-rate assumption ($\nabla \cdot \boldsymbol{v}_s = 0$) is used. 

In the coordinate system where the $x$-axis is parallel to and the $z$-axis is normal to the slab extension, if the angle of flow direction ($\theta$) is uniform across the slab (Fig. \ref{fig: geometry}b), then $\boldsymbol{v}_s$ and $\boldsymbol{v}_f$ in 2D can be written as:
\begin{equation} \label{eq: velvec}
\begin{split}
\boldsymbol{v}_s &= (v_s, \: 0), \\
\boldsymbol{v}_f &= v_f(\cos \theta, \: \sin \theta),
\end{split}
\end{equation}
where $v_s$ is the subduction rate, $v_f$ is the magnitude of fluid velocity, and $\theta$ represents flow direction, a model parameter that will be explored in section \ref{sec: results}. Substituting equation~\eqref{eq: velvec} into \eqref{eq: bulkmass} and \eqref{eq: bulkspc} yields:
\begin{equation} \label{eq: gov1}
-(\rho_s - \rho_f)\frac{\partial \phi}{\partial t} + \rho_f\left[ \cos \theta \frac{\partial (v_f \phi)}{\partial x} + \sin \theta \frac{\partial (v_f \phi)}{\partial z} \right] - \rho_s v_s \frac{\partial \phi}{\partial x} = 0,
\end{equation}
\begin{equation} \label{eq: gov2}
\frac{\partial \overline{c^i}}{\partial t} + \rho_f \cos \theta \frac{\partial (v_f \phi c^i_f)}{\partial x} + \rho_f \sin \theta \frac{\partial (v_f \phi c^i_f)}{\partial z}  +  \rho_s v_s \frac{\partial}{\partial x} \left[ (1 - \phi) c^i_s \right] = 0,
\end{equation}
where $\overline{c^i}$ is volatile mass per unit rock volume and can be converted to mass fraction via:
\begin{equation} \label{eq: gov3}
\overline{c^i_{th}} = \frac{\overline{c^i}}{\rho_f \phi + \rho_s (1 - \phi)},
\end{equation}
which is used as input for the thermodynamic module from the companion paper Part I (the subscript ``$th$" indicates input variables for the thermodynamic module):
\begin{equation} \label{eq: gov4}
\left( \phi, \: c^{H_2O}_s, \: c^{H_2O}_f, \: c^{CO_2}_s, \: c^{CO_2}_f \right) = \Phi \left( P, \: T, \: \overline{c^{H_2O}_{th}}, \: \overline{c^{CO_2}_{th}} \right).
\end{equation}
Equations~\eqref{eq: gov1}--\eqref{eq: gov4} constitute the governing equations in this study. 

\begin{table}
\caption{Notations of Symbols}
\label{tab: symbol}
\centering
\begin{ruledtabular}
\begin{tabular}{l c r} 

\textbf{Symbol}  & \textbf{Meaning} & \textbf{Unit} \\
\hline
   $D$  & diffusion coefficient of $\mathrm{CO_2}$ or $\mathrm{H_2O}$  &  $\sim$$10^{-8}$ $\mathrm{m / s^{2}}$  \\
   $P$  & pressure  &  GPa  \\
   $T$  & temperature  &  $^{\circ}$C  \\
   $H$  & slab thickness  &  m  \\
   $i$  &   $\mathrm{H_2O}$ or $\mathrm{CO_2}$ \\
   $t$  & time  &  s \\
   $x$  & coordinate along slab  &  m \\
   $z$  & coordinate normal to slab  &  m \\
   $\theta$  & uniform flow angle   \\
   $\phi$  & porosity   \\
   $\boldsymbol{v}_f$  & fluid velocity  &  $\mathrm{m / s}$   \\
   $v_f$  & magnitude of fluid velocity   &  $\mathrm{m / s}$  \\
   $\boldsymbol{v}_s$  & solid velocity  &  $\mathrm{m / s}$  \\
   $v_s$  & subduction rate  &  $\mathrm{m / s}$  \\
   $\rho_f$  & fluid phase density  &  $\mathrm{kg / m^{3}}$ \\
   $\rho_s$  & solid phase density  &  $\mathrm{kg / m^{3}}$ \\
   $c^{i}_f$  & mass fraction of $i$ in fluid phase   \\
   $c^{i}_s$  & mass fraction of $i$ in solid phase   \\
   $\overline{c^i}$  & bulk content of $i$ per unit rock volume  &  $\mathrm{kg / m^{3}}$ \\
   $\overline{c^i_{th}}$  & bulk mass fraction of $i$   \\
   $\Gamma$  & reaction rate  &  $\mathrm{kg / m^{3} / s}$  \\
   $\Gamma_i$  & reaction rate for $i$  &  $\mathrm{kg / m^{3} / s}$  \\

\end{tabular}
\end{ruledtabular}
\end{table}

The unknowns in the governing equations are the magnitude of fluid velocity ($v_f$), porosity ($\phi$), bulk compositions ($\overline{c^i}$), and $\mathrm{H_2O}$ \& $\mathrm{CO_2}$ content in both phases ($c^i_f$, $c^i_s$)---8 unknowns in total.  The thermodynamic equation \eqref{eq: gov4} provides solutions to 5 unknowns, so it can be counted as 5 equations. Including equations \eqref{eq: gov1}--\eqref{eq: gov2}, the total number of equations is 8, so the set of governing equations is closed. Note that the closedness of governing equations is premised on the prescription of flow angle ($\theta$) and steady-state $P$--$T$ structure of the slab. To be consistent with the steady-state $P$--$T$ field, we solve the governing equations~\eqref{eq: gov1}--\eqref{eq: gov4} until a steady state is reached (after $\sim$6 Ma model time for a 5~cm~yr$^{-1}$ convergence rate). 

The model yields the magnitude distribution of fluid velocity with pre-defined uniform flow direction by mass conservation, without resorting to the momentum conservation equation (i.e., Darcy's law). With the $P$--$T$ path of a subducting slab pre-determined, thermodynamics of local equilibrium dictates the amount of volatiles liberated from or absorbed to every rock parcel during each timestep of the model. Under steady state, with flow direction ($\theta$) prescribed, mass conservation dictates that the magnitude of fluid flux is simply the integration of volatiles released or absorbed (the term $v_s (\partial \phi / \partial x)$ in equation~\eqref{eq: gov1}) along flow trajectories. Readers are referred to Appendix~\ref{apx:a} for further details.

We note that avoiding the momentum conservation equation by prescribing the flow direction ($\theta$) has positive and negative consequences. The downside is the lack of emergent, dynamic flows; instead, fluid flows are determined by choice of flow direction and conservation of mass along the selected flow paths. As briefly discussed in the introduction, the upside is that the large uncertainties associated with the diversity of possible flow patterns yielded by various dynamic considerations is avoided. For example, two-phase dynamic model using viscous rheology suggests a high-permeability flow channel at the interface between slab and mantle wedge \citep{Wilson:2014aa}, whereas those using visco-elastic rheology show the development of porosity waves within the slab \citep{Morishige:2018aa}. On smaller spatial scales (e.g., centimeter to meter), \citet{Plumper:2017aa} and \citet{Malvoisin:2015aa} show that flow channels and porosity waves can develop as well. Besides, visco-elasto-plastic models suggest that faults formed during slab bending exert a strong control on the fluid flow directions and overall flow pattern within the slab \citep{Faccenda:2009aa}. Indeed, there is various field evidence attesting to the complexity of fluid flows in subducting slabs \citep[e.g.,][]{Ague:2007aa, Philippot:1991aa, Barnicoat:1995aa, Putlitz:2000aa, Breeding:2003aa, Breeding:2004aa, Galvez:2013aa, Piccoli:2016aa, Piccoli:2018aa}. Designating flow direction ($\theta$) as a free model parameter, however, enables us to approximate the overall trend of fluid flows and explore it by setting $\theta$ to different values. Nevertheless, ignoring the flow dynamics does not revert the current model to the conventional open-system models. For example, the fraction of exsolved volatiles that is moved out and reacted upwards is determined ad hoc \citep[e.g.,][]{Gorman:2006aa}, but this fraction is constrained by mass conservation in the current model such that porosity is maintained at a finite steady level.

In summary, the assumptions regarding the thermo-mechanical aspects of the slab in our model are: (i) rigid plate with a prescribed subduction rate $v_s$; (ii) constant densities for liquid ($\rho_f$) and solid ($\rho_s$) phases; (iii) steady-state $P$--$T$ structure of the slab pre-calculated from canonical thermo-mechanical models. The assumptions in the reactive flow model are: (i) local equilibrium between liquid and solid phases; (ii) no diffusion or dispersion of H$_2$O and CO$_2$; (iii) flow direction $\theta$ is prescribed and uniform across the slab; (iv) the reactive flow is solved until a steady state is reached.

\section{Numerical Method} \label{method}
\subsection{Initial and Boundary Conditions}
We undertake timestepping for equations~\eqref{eq: gov1}--\eqref{eq: gov4} to achieve the steady-state solution. The initial conditions are: bulk volatile content $\overline{c^{i}_{th}}$ are uniformly set to those of the incoming rocks on the left boundary; porosity ($\phi$) and volatile content in both phases ($c^{i}_{f}$, $c^{i}_{s}$) are determined via the thermodynamic module (eq.~\eqref{eq: gov4}); initial fluid speed ($v_f$) is zero across the modelled slab domain.

The incoming slab is assumed to be layered in lithology (Fig. \ref{fig: geometry}b) and thus has bulk volatile content: $c^{\mathrm{CO_2}}_{th} = 3.01$ wt\% and $c^{\mathrm{H_2O}}_{th} = 7.29$ wt\% \citep{Plank:1998aa} for a sedimentary layer from 0 km to 0.5 km; $c^{\mathrm{CO_2}}_{th} = 2.95$ wt\% and $c^{\mathrm{H_2O}}_{th} = 2.68$ wt\% \citep{Kerrick:2001ab} for a MORB layer from 0.5 km to 2.5 km; $c^{\mathrm{CO_2}}_{th} = 2.84$ wt\% and $c^{\mathrm{H_2O}}_{th} = 2.58$ wt\% for a gabbroic layer from 2.5 km to 6.5 km; and $c^{\mathrm{CO_2}}_{th} = 0.02$ wt\% and $c^{\mathrm{H_2O}}_{th} = 1$ wt\% \citep{Hart:1986aa} for a slab-mantle layer from 6.5 km to 8 km. Note that $1$ wt\% H$_2$O content in serpentinized peridotite roughly corresponds to a serpentinization degree of $\sim$8\% \citep{Carlson:2003aa}. The H$_2$O and CO$_2$ content in the altered gabbro is adopted from typical metabasalt values \citep{Kerrick:2001ab}, and re-normalized using the non-volatile composition for gabbro in \citet{Hacker:2008aa}. This very likely represents an upper limit because hydrothermal alteration decreases with depth \citep{Kelemen:2015aa}. In fact, gabbros altered in mid-ocean ridges are estimated to contain 0.2--1.3 wt\% H$_2$O \citep{Hacker:2003aa, Carlson:2003ab}, but all these estimates exclude the contribution from volatile addition in the outer rise near subduction \citep{Peacock:2001aa}. If hydrothermal changes at both mid-ocean ridges and outer rises are taken into account, the high gabbro volatile content adopted here could represent an upper limit.

As for boundary conditions, since the governing equation for flow speed ($v_f$) is first order in space, only one boundary condition along the flow paths is needed. For flow paths originating from the slab base, no-flux ($v_f=0$) boundary condition is prescribed (Fig. \ref{fig: geometry}b). However, for the ghost points above the upper boundary in the numerical grid, Neumann boundary conditions ${\partial v_f}/{\partial z}$ and ${\partial \overline{c^i_{th}}}/{\partial z}$ are adopted for flow speed and bulk volatile content. The left and right boundary conditions depend on flow direction ($\theta$). When $\theta \leq 90 ^{\circ}$, the flow paths originate from the left and basal boundaries, so we let $v_f =0 $ and $\overline{c^i_{th}}=\overline{c^i_{th}} \Big|_{\mathrm{incoming}}$ on the left, and ${\partial v_f}/{\partial x} = 0$ and ${\partial \overline{c^i_{th}}}/{\partial x} = 0$ on the right. When $\theta > 90 ^{\circ}$, the flow paths originate from the right and basal boundaries, so we let ${\partial v_f}/{\partial x} = 0$ and $\overline{c^i_{th}}=\overline{c^i_{th}} \Big|_{\mathrm{incoming}}$ on the left, and $v_f =0 $ and ${\partial \overline{c^i_{th}}}/{\partial x} = 0$ on the right.

\subsection{Solution Procedure}
The nonlinear character of the governing equations~\eqref{eq: gov1}--\eqref{eq: gov4} is evident by inspection. For example, equation \eqref{eq: gov3} shows that bulk volatile content depends on fluid flow ($v_f, \theta$), porosity ($\phi$), and volatile content in each phase ($c^i_f$, $c^i_s$), among which porosity and phase volatile content depend further on bulk compositions through thermodynamics expressed in equation \eqref{eq: gov4}. To solve this equation set, we employ PETSc \citep[Portable, Extensible Toolkit for Scientific computing,][]{petsc-web-page}, following the procedure illustrated in \citet{Katz:2007aa} to write down the residuals of governing equations~\eqref{eq: gov1}--\eqref{eq: gov2}. Every timestepping solve is handled by the nonlinear solver provided by PETSc, and the special treatment in our solution procedure is that we emplace the thermodynamic module (eq.~\eqref{eq: gov4}) within every evaluation of residuals at each Newton iteration.

At the sites where devolatilization onsets, there will be pulses of volatile production owing to the increase of porosity from ideally zero to some finite value determined by thermodynamics. These pulses of volatile production along the envelope of the devolatilizing region eventually contribute to integrated volatile fluxes, and cause small fluctuations in the computed fluxes. To focus on the general trend and overall pattern of volatile fluxes within and atop the slab, the results returned by the solution procedure above are post-processed through a smoothing step. Different length scales of smoothing have been experimented with and we choose 12 numerical grid points (corresponds to 3 km) as the spatial span for smoothing. Details on the smoothing are provided in Appendix \ref{apx:a}.

\section{Results} \label{sec: results}

Before presenting the results for the open-system reference model, we first provide the closed-system results (Fig.~\ref{fig: 4layer-closed}) with a two-fold goal: firstly to compare our closed-system results with the previous, purely thermodynamic model that assumes a closed system, and secondly for later comparison with the open-system reference model to elucidate the open-system effects. 

\subsection{Closed System} \label{sec: closedmodel}
The closed system is characterized by the fact that there are no mass fluxes into or out of the rocks of interest, leaving the bulk compositions unaltered. As a result, the bulk $\mathrm{CO_2}$ \& $\mathrm{H_2O}$ content of the incoming rocks stay unchanged as subduction proceeds with a speed $v_s$; that is, these bulk values in the slab remain equal to those at the inflow boundary (left side in Fig.~\ref{fig: geometry}b). Such a closed-system model corresponds to that typically used in earlier studies, where a representative but fixed bulk composition is used for each rock type to construct thermodynamic phase diagrams over which various subduction geotherms are superimposed \citep{Peacock:1990aa, Peacock:1991aa, Hacker:2003aa, Keken:2011aa, Hacker:2008aa}.

\begin{figure}[h!]
\centering
\includegraphics[scale=0.45]{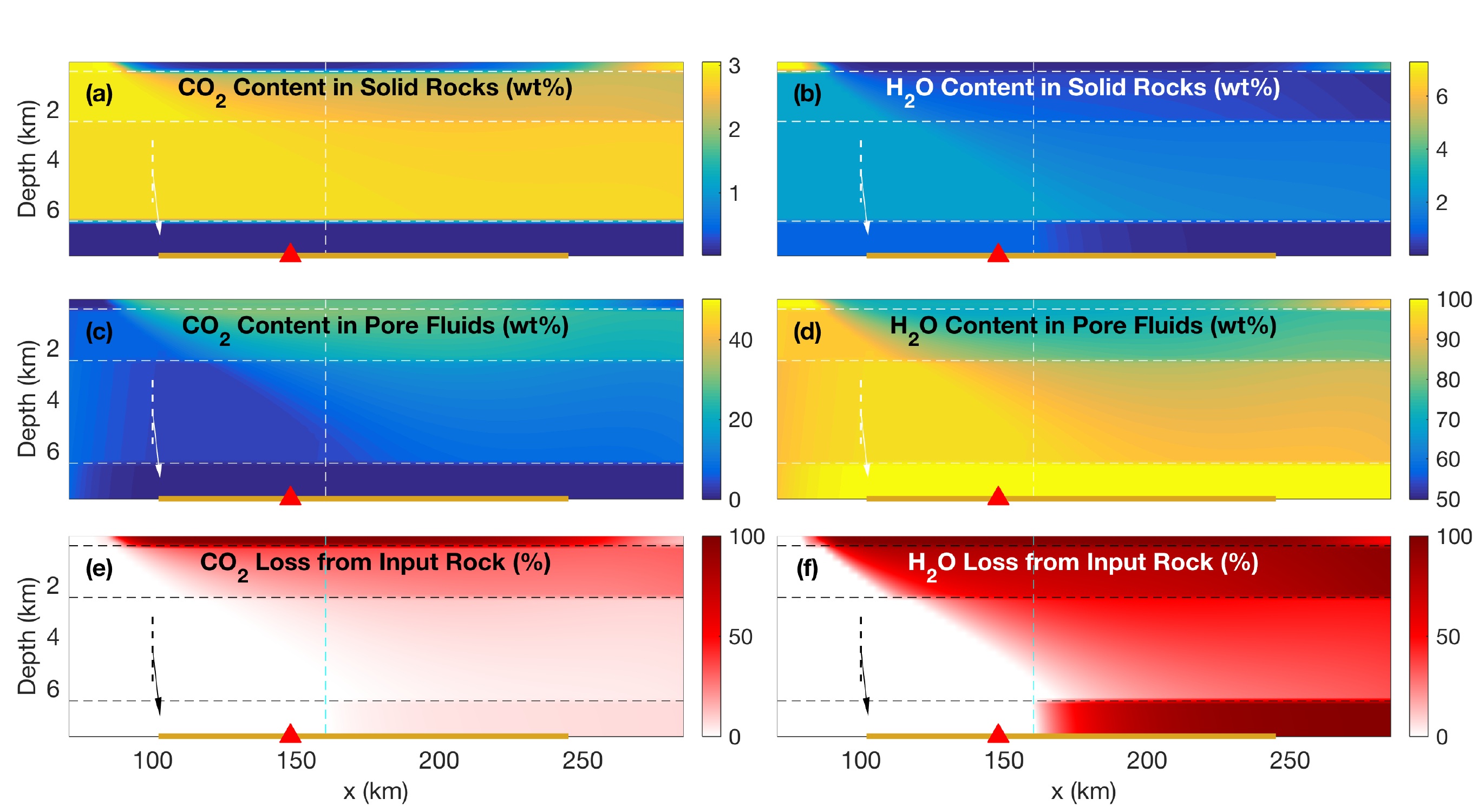}
\caption[Optional caption for list of sub figures]{Results for the closed-system model. Horizontal dashed lines (white or black) mark the lithological interfaces, and long vertical dashed lines mark the position where the basal serpentinized upper mantle starts to devolatilize. Divergent colormaps (blue-white-red) are used for panels (e) and (f) where red color corresponds to positive values (volatile loss) and blue color corresponds to negative values (volatile gain). Volatile loss (or gain) in (e) and (f) is calculated relative to its content on the left boundary. Succeeding figures adopt the same line, arrow and color conventions. The brown line and red triangle at the base of each panel have the same meaning as in Figure \ref{fig: ptstruct}, but are placed at the panel base to avoid cluttering the display of the top sedimentary layer. }
\label{fig: 4layer-closed}
\end{figure}

With the $P$--$T$ field of a hot slab (Fig. \ref{fig: ptstruct}), Figure~\ref{fig: 4layer-closed} shows our closed-system results. Left panels are results on CO$_2$, and right panels on H$_2$O. From top to bottom, the rows of panels show CO$_2$ and H$_2$O content in the solid rock phase, liquid volatile phase, and their loss relative to the starting values on the inputting boundary. Of course in a closed system, there should be no volatile loss. However, in closed-system devolatization models it is conventionally assumed that volatiles, once exsolved, are immediately extracted from the bulk system (without altering the bulk composition) \citep[e.g.,][]{Peacock:1990aa, Peacock:1991aa, Hacker:2003aa, Keken:2011aa, Hacker:2008aa}. In this context, a general trend of CO$_2$ and H$_2$O loss from rocks can be seen with along-slab distance in Figure \ref{fig: 4layer-closed}a--b, and more clearly in Figure \ref{fig: 4layer-closed}e--f. However, there is a reversal of this trend near the slab distance $\sim$210 km in the top sedimentary layer (e.g., Fig. \ref{fig: 4layer-closed}a). It can also be seen from Figure \ref{fig: 4layer-closed}c--d that, within each lithological layer, the coexisting liquid phase is gradually more CO$_2$-rich as subduction goes deeper, except where the trend reversal takes place. 

Under the model assumption regarding extraction from a closed system, the volatile loss in Figure \ref{fig: 4layer-closed}e--f is the change in volatile content of the solid phase from its initial value. Since the amount of liquid phase in equilibrium depends on $P$--$T$ conditions, the non-uniform $P$--$T$ field within the slab yields varying extent of volatile loss in the closed-system model. The trend reversal in Figure~\ref{fig: 4layer-closed} can be understood by superimposing the top-layer geotherms over the phase diagram for closed systems \citep[e.g.,][]{Keken:2011aa}. The $P$--$T$ conditions in the segment beyond $\sim$210 km in the top layer exceed the $P$--$T$ curve for the onset of devolatilization, but to a lesser extent than the $P$--$T$ conditions in the segment before $\sim$210 km, leading to a smaller degree of devolatlization. Similar trend reversal of closed-system volatile loss can also be observed for the warm subduction geotherm in \citet[][figure 3a]{Kerrick:2001aa}.

Depths of volcanic arcs to slab surface range from $\sim$70 km to $\sim$170 km \citep{Syracuse:2006aa}, which correspond to along-slab distances of $\sim$100 km to $\sim$250 km in Figure \ref{fig: 4layer-closed}.  Closed-system model predictions, shown in Figure \ref{fig: 4layer-closed}a--b and e--f, indicate that slab sediments almost completely lose $\mathrm{CO_2}$ \& $\mathrm{H_2O}$ at forearc depths; the basaltic and gabbroic layers can supply significant $\mathrm{H_2O}$ but only limited $\mathrm{CO_2}$ at subarc depths; the slab-mantle lithosphere can release almost all of its $\mathrm{H_2O}$ at subarc depths. All these results are consistent with the previous studies that assume closed systems and consider hot subduction geotherms \citep{Kerrick:1998aa, Kerrick:2001aa, Kerrick:2001ab}.

To facilitate later comparison, it is worth noting the implications of the closed-system model for the open-system behaviors in the succeeding sections. Firstly, the basal slab-mantle layer in this model contains negligible CO$_2$ and serves essentially as a water supplier. Because hydrated slab mantle has only a small set of hydrous minerals (i.e., talc, brucite, serpentine, chlorite), H$_2$O loss in this layer is more abrupt and complete relative to in basaltic and sedimentary layers. A sudden onset of the supply of the basal H$_2$O will, in later sections on open-system flow, lead to significant H$_2$O infiltration. Secondly, at the peridotite--gabbro and gabbro--basalt interfaces (6.5 and 2.5~km deep into the slab, respectively), there is a sharp CO$_2$ concentration gradient in the liquid phase (Fig.~\ref{fig: 4layer-closed}c--d). Fluid ascent in the open systems would inevitably cause H$_2$O-rich and CO$_2$-poor fluid infiltration that enhances decarbonation \citep{Gorman:2006aa}. Thirdly, the CO$_2$ concentration in the liquid phase decreases from the basaltic to the sedimentary layer, suggesting the potential for carbonation by fluid flow down the gradient of CO$_2$ concentration. 

\subsection{Open-System Reference Model} \label{sec: refmodel}
\begin{figure}[h!]
\centering
\includegraphics[scale=0.45]{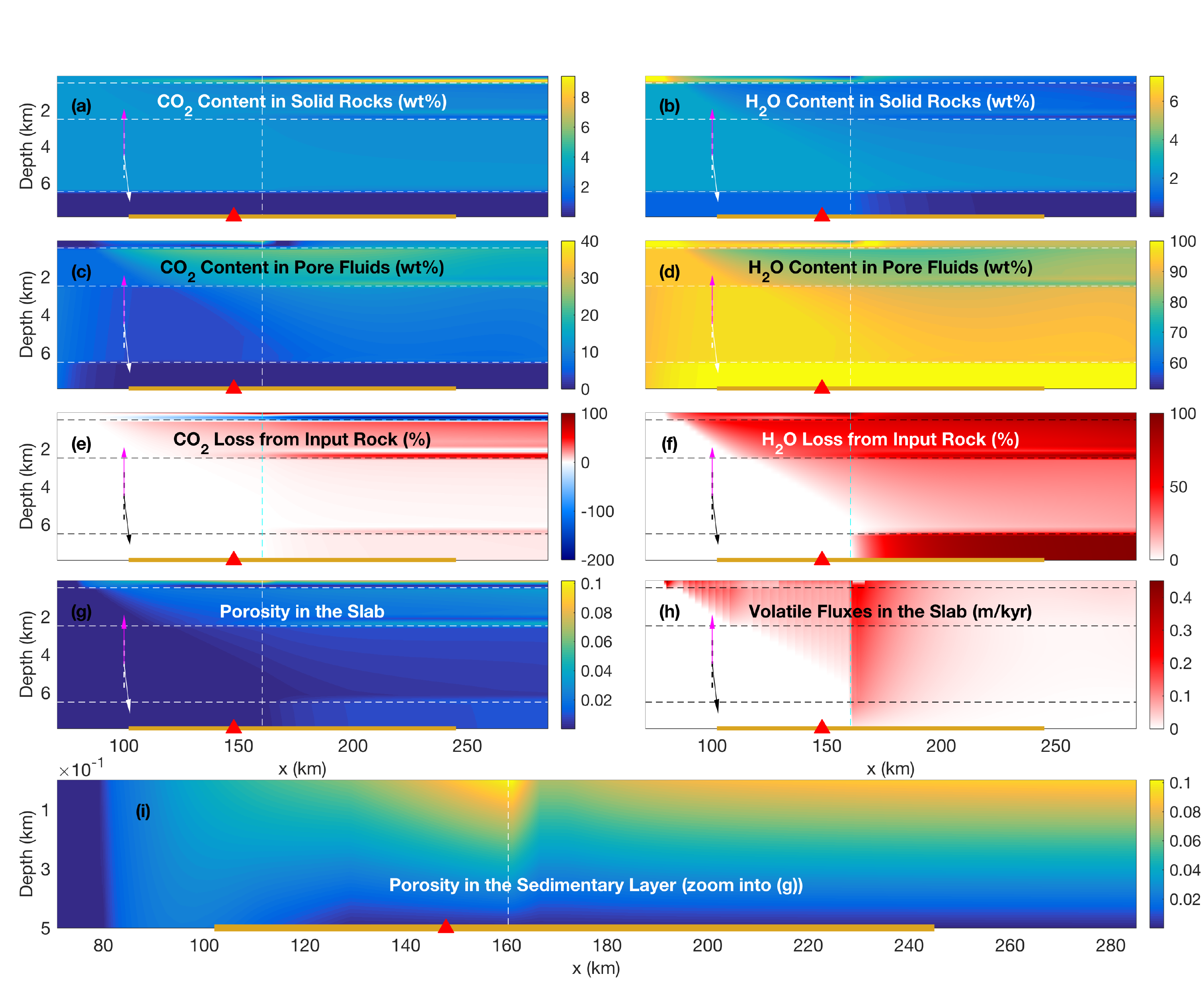}
\caption[Optional caption for list of sub figures]{Results for the open-system reference model. The magenta arrows denote the direction of fluid flow (90$^{\circ}$ in the reference model) relative to slab extension, whereas the black or white arrows denote the direction of gravity. The short dashed lines (white or black) are a reference direction normal to the slab extension. Red color in the divergent colormap for volatile flux (panel h) means upward fluid flow, whereas blue color means downward flow. Note the different horizontal and vertical scales in the plots, and the true angle between gravity and the flow direction should be 145$^{\circ}$. Panel (i) maps the porosity distribution only for the top sedimentary layer in panel (g). }
\label{fig: refmodel}
\end{figure}

\begin{figure}[h!]
\centering
\includegraphics[scale=0.5]{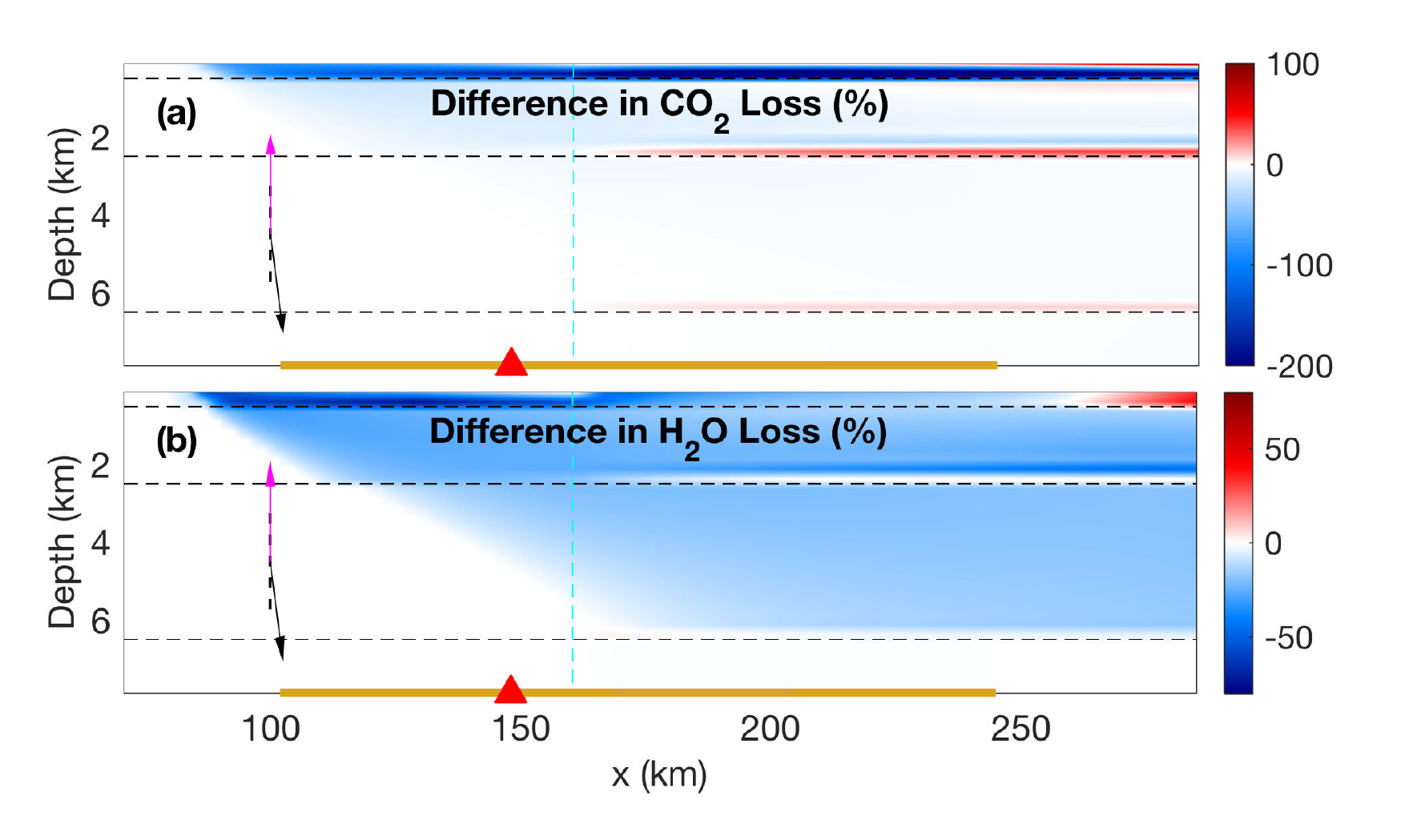}
\caption[Optional caption for list of sub figures]{Differences in $\mathrm{CO_2}$ and $\mathrm{H_2O}$ loss (or gain) between the open- and closed-system models. The difference is calculated by rock $\mathrm{CO_2}$ or $\mathrm{H_2O}$ loss in the open system minus that in the closed system, and percentage values are relative to the initial rock volatile content. Note that, different than those in Figures \ref{fig: 4layer-closed} and \ref{fig: refmodel}, red color means enhanced loss or diminished gain, whereas blue color means diminished loss or enhanced gain.}
\label{fig: diffopenclose}
\end{figure}

Taking all the parameters for the closed-system model above, the reference model additionally considers fluid flows with a uniform direction $\theta = 90 ^{\circ}$ (Fig.~\ref{fig: geometry}b), namely, fluid flow is normal to the direction of slab motion ($\boldsymbol{v}_s$). The results of the steady-state solution to this open-system model are presented in Figure \ref{fig: refmodel}. The panel layout of Figure \ref{fig: refmodel} is the same as that of Figure \ref{fig: 4layer-closed} except for the two additional panels in the bottom row. These show the porosity (Fig.~\ref{fig: refmodel}g) and volatile flux (Fig.~\ref{fig: refmodel}h) distribution within the slab. To elucidate and further highlight the open-system effects, Figure \ref{fig: diffopenclose} maps the differences in CO$_2$ and H$_2$O loss (or gain) between the closed- and open-system models, that is, volatile loss in the open system minus that in the closed system. We next consider the reference model results in each layer of the slab, from bottom to top.

The basal slab-mantle layer does not experience fluid infiltration because the flows are upward and there is no flux at the bottom boundary. Figure \ref{fig: refmodel}c--d shows that the devolatilized liquid phase is almost pure H$_2$O, given that the inputting peridotitic rock contains negligible bulk CO$_2$ (0.02 wt\%). The H$_2$O loss is thus abrupt and complete (Fig. \ref{fig: refmodel}f), similar to that for the closed-system model. 

On top of the base layer, the gabbro layer is fluxed by nearly pure H$_2$O sourced from below (Fig. \ref{fig: refmodel}c--d). As a result, the infiltration and fractionation effects discussed in Part I come into play. The very H$_2$O-rich and CO$_2$-poor fluids will infiltrate the base of the gabbro layer, causing enhanced decarbonation and inhibited dehydration. The enhanced CO$_2$ loss is evident in Figure \ref{fig: diffopenclose}a, but the expected reduction in H$_2$O loss is not as clear in Figure \ref{fig: diffopenclose}b. This can be explained by the fractionation effect detailed in Part I. Because increased decarbonation reduces the ratio of CO$_2$ over H$_2$O in the bulk system of solid plus liquid, the onset temperature of devolatilization decreases at the base of the gabbro layer. Given that temperatures within the modelled slab are held fixed, such a decrease in the onset temperature means devolatilization of the bulk system becomes easier, so the resultant increase of devolatilization extent can offset the inhibitive effect on H$_2$O loss induced by the infiltration. The elevated extent of devolatilization is also reflected by the higher porosity at the base of the gabbro layer in Figure \ref{fig: refmodel}g. 

The infiltration effect is strongest at the peridotite--gabbro interface because the concentration gradients of $\mathrm{CO_2}$ and $\mathrm{H_2O}$ in the liquid phase are the steepest (Fig. \ref{fig: refmodel}c--d). Above the interface, the liquid phase compositions have been adjusted by the buffering reactions drastically occurring at the interface, so the infiltration effect weakens. In consequence, the bulk interior of the gabbro layer is dominated by the fractionation effect. At and after the onset of devolatilization, preferential $\mathrm{H_2O}$ loss causes increased difficulty in devolatilization and thus inhibits overall volatile loss in the bulk interior (Fig. \ref{fig: diffopenclose}).

Further up, the H$_2$O-rich and CO$_2$-poor fluids (Fig. \ref{fig: refmodel}c--d) derived from the gabbro layer infiltrate the base of the basalt layer, similar to the scenario at the base of gabbro layer. Likewise, relative to the closed system, enhanced $\mathrm{CO_2}$ loss and approximately unmodified $\mathrm{H_2O}$ loss take place at the layer base (Fig. \ref{fig: diffopenclose}), and the overall $\mathrm{CO_2}$ and $\mathrm{H_2O}$ loss is inhibited in the layer interior. The porosity elevation immediately above the gabbro--basalt interface is even more discernible in this case (Fig. \ref{fig: refmodel}g). Moreover, elevated porosity levels reflect higher extent of devolatilization. According to our parameterization in the companion paper Part I, higher extent of devolatilization corresponds to higher CO$_2$ and lower H$_2$O concentrations in the coexisting liquid phase, which are also evident at $\sim$2.5 km depth in Figure \ref{fig: refmodel}c--d. Upward infiltration of these fluids, coupled with the fractionation effect, gives rise to the diminished volatile loss immediately above the region of enhanced decarbonation (blue stripes $\sim$2 km deep in Fig. \ref{fig: diffopenclose}). Note also that in Figure \ref{fig: diffopenclose} the degree of decarbonation is higher at the basalt base than at the gabbro base, which is consistent with the higher integrated fluid fluxes at the shallow lithological interface (Fig.~\ref{fig: refmodel}h).

Figure \ref{fig: refmodel}c--d shows that the fluid infiltration at the base of the topmost sedimentary layer is down the CO$_2$ and up the H$_2$O concentration gradients. As is inferred from the closed-system model, the infiltration will result in carbonation reactions in the sediments immediately above the lithological contact, which is demonstrated by the negative values of CO$_2$ loss (i.e., CO$_2$ gain by the sediments) in Figure \ref{fig: refmodel}e. Figure \ref{fig: diffopenclose}a compares the CO$_2$ loss between the open- and closed-system models; carbon sequestration in the top sedimentary layer is more evident. In Figure \ref{fig: diffopenclose}b, it can be seen that H$_2$O loss in the sedimentary layer diminishes before $\sim$250 km along the slab coordinate. This is due to the fractionation effect discussed in Part I: carbonation increases the ratio of bulk CO$_2$ over H$_2$O, which further raises the onset temperature of devolatilization and in turn inhibits dehydration. As for the enhanced H$_2$O loss after $\sim$250 km in Figure \ref{fig: diffopenclose}b, it actually stems from the reversal of trend in the H$_2$O loss calculated for the closed-system model (e.g., Fig.~\ref{fig: 4layer-closed}f).

A common feature above the gabbro--basalt and basalt--sediment interfaces in Figure \ref{fig: diffopenclose}a is the formation of paired stripes in red and blue, but the vertically reversed color order for the stripe pair in the sedimentary layer indicates that decarbonation follows carbonation along infiltration. This is because the former stripe pair is caused by infiltration down a $\mathrm{CO_2}$ gradient, whereas the latter is by that up the gradient. Another noteworthy feature in Figure \ref{fig: diffopenclose} is that the starting positions of the infiltration-induced stripes in the basalt and gabbro layers coincide horizontally with the onset of dehydration of the slab upper mantle. This coincidence reflects the potential for fluids from hydrated lithospheric mantle to remobilize slab H$_2$O and CO$_2$. In fact, the enhanced infiltration into the top sedimentary layer causes considerable carbonation that substantially reduces the porosities and fluid fluxes immediately after $\sim$160 km along the slab (Fig.~\ref{fig: refmodel}g--i).

Close examination of Figure \ref{fig: refmodel}h suggests that the volatile fluxes fluctuate laterally before the onset of slab mantle dehydration at $\sim$160 km. This is because the onset of devolatilization at different slab depths occurs at different horizontal positions that are laterally separated from one another. Since the onset of devolatilization corresponds to a pulse of volatile production, after being vertically integrated into volatile fluxes, these pulses cause lateral fluctuation in volatile flux distribution. Appendix \ref{apx:a} discusses this issue further. 

\subsection{Effect of Flow Direction} \label{sec: flowdirection}
\begin{figure}[h!]
\centering
\includegraphics[scale=0.45]{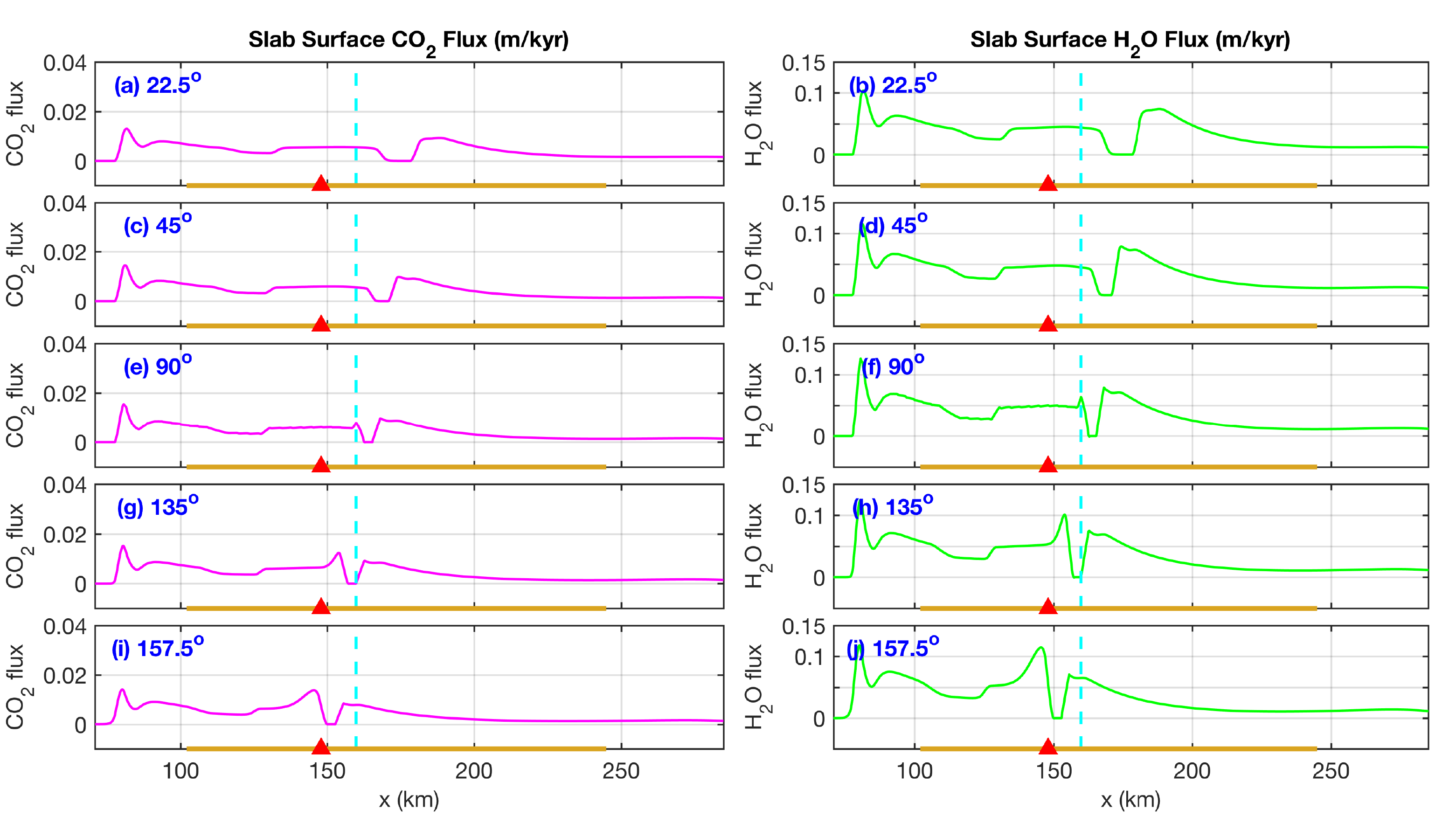}
\caption{$\mathrm{CO_2}$ (left panels) and $\mathrm{H_2O}$ (right panels) fluxes along the slab surface as a function of fluid flow direction ($\theta$). Note that the panels (e and f) in the middle row are the fluxes from the reference model in section \ref{sec: refmodel}. Cyan dashed lines mark the starting position of slab mantle devolatilization and this convention applies in the succeeding figures. }
\label{fig: surfacefluxcomp}
\end{figure}

Different flow trajectories pass through areas with different $P$--$T$ conditions that are associated with different volatile partition coefficients. Hence volatile loss and gain during reactive flow are path-dependent, and we assess the impact of fluid flow direction ($\theta$) on CO$_2$ and H$_2$O fluxes emanating from slab surface. Figure \ref{fig: surfacefluxcomp} shows the slab surface CO$_2$ (left panels) and H$_2$O (right panels) fluxes with flow angles of $22.5^{\circ}$, $45^{\circ}$, $90^{\circ}$, $135^{\circ}$, and $157.5^{\circ}$. The case of flow angle $\theta = 90^{\circ}$ is the open-system reference model already presented in section \ref{sec: refmodel}. Other than flow angle, all the parameters for model runs in Figure \ref{fig: surfacefluxcomp} are identical to those in the reference model. The flow angles $22.5^{\circ}$, $45^{\circ}$, $90^{\circ}$, $135^{\circ}$, and $157.5^{\circ}$ are chosen to spread between $0^{\circ}$ and $180^{\circ}$, and thus representative of various flow scenarios; other choices of flow angles don't affect our conclusion. We also note that the angle of $22.5^{\circ}$ corresponds to downward flow, which could occasionally occur in localized settings \citep[e.g.,][]{Faccenda:2009aa}. The consideration of this flow scenario is primarily to explore as completely as possible the spectrum of within-slab flow patterns.

Figure \ref{fig: surfacefluxcomp} shows profiles of the slab surface volatile flux (i.e., $[v_f \phi c_f^i]_{z=0} \sin \theta$). All the profiles have a peak near $\sim$80 km along the slab, caused by the devolatilization of the top sedimentary layer. In addition, there is a dip in all the flux profiles around the starting position of slab-mantle devolatilization ($\sim$160 km). This sharp drop of surface flux stems from the boost of carbonation reactions by slab-mantle devolatilization, which reduces the top-layer porosity ($\phi$) and further the surface flux ($v_f \phi$). The shift in position of this dip is due to the change of flow direction that advances ($\theta > 90^{\circ}$) or retreats ($\theta < 90^{\circ}$) the carbonation reaction in the top sedimentary layer. A more striking feature in Figure \ref{fig: surfacefluxcomp} is a second flux peak that emerges at the average position of global arcs projected onto the slab surface (red triangle). As shown in Figure \ref{fig: surfacefluxcomp}g--j, this peak appears only when flow angle $\theta > 90^{\circ}$, that is, for fluid flows that are nearly slab-parallel and up-dip. If it is further assumed that magmas generated by flux melting traverse the mantle wedge vertically without deflection \citep[e.g.,][]{Grove:2012aa}, Figure \ref{fig: surfacefluxcomp} implies that up-slab fluid flow is responsible for the magma supply to volcanic arcs.

\begin{figure}[h!]
\centering
\includegraphics[scale=0.45]{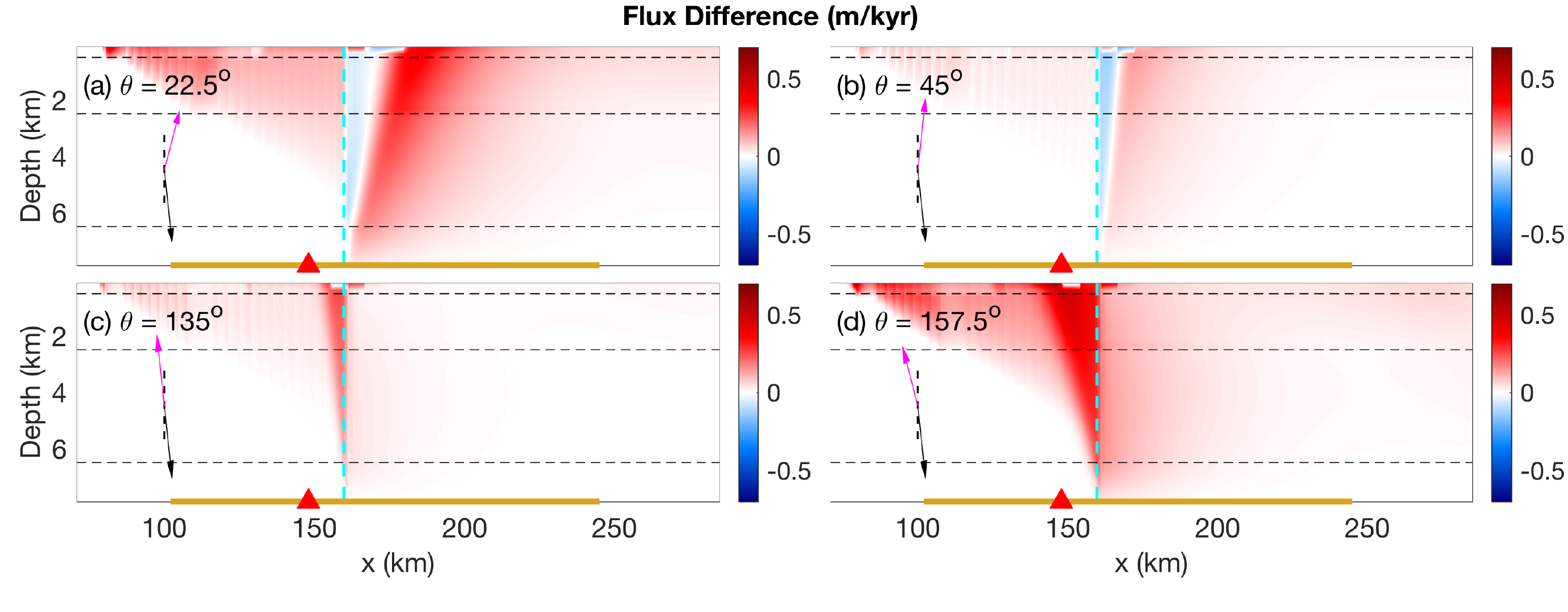}
\caption{The difference in fluid fluxes ($v_f \phi$) between the models with flow angle $\theta \neq 90^{\circ}$ and the open-system reference model with $\theta = 90^{\circ}$. Red color corresponds to flux increase, and blue color to flux decrease, both relative to the reference model results. Arrows and dashed lines bear the same meaning as in preceding figures. Note that the small patches in the sedimentary layer around $\sim$160 km is related to the sharp porosity reduction due to strong carbonation, read the reference model section \ref{sec: refmodel} for more details. }
\label{fig: dir-flux}
\end{figure}

The emergence of the subarc flux peak in Figure \ref{fig: surfacefluxcomp} can be explained by Figure \ref{fig: dir-flux}, which maps the difference in fluxes ($v_f \phi$) between models with flow angle $\theta \neq 90^{\circ}$ and the reference model with $\theta = 90^{\circ}$. The flux increase in Figure \ref{fig: dir-flux} can be understood by inspecting equation~\eqref{eq: fluxint}: fluid fluxes accumulate along flow paths and longer paths tend to cause larger fluxes. Moreover, when $\theta > 90^{\circ}$, additional flux increase is caused by flow trajectories tapping into the region of slab-mantle devolatilization (Fig. \ref{fig: dir-flux}c--d). Conversely, when $\theta < 90^{\circ}$, the flux diminution in Figure \ref{fig: dir-flux}a--b is caused by flow trajectories passing through regions before the onset of slab mantle devolatilization. As flows with $\theta > 90^{\circ}$ can transport more volatiles to slab surface, this explains the emergence of flux peak near $\sim$150 km in the flux profiles in Figure \ref{fig: surfacefluxcomp}g--j. 

\subsection{Effect of Slab Age} \label{sec: age}
\begin{figure}[h!]
\centering
\includegraphics[scale=0.45]{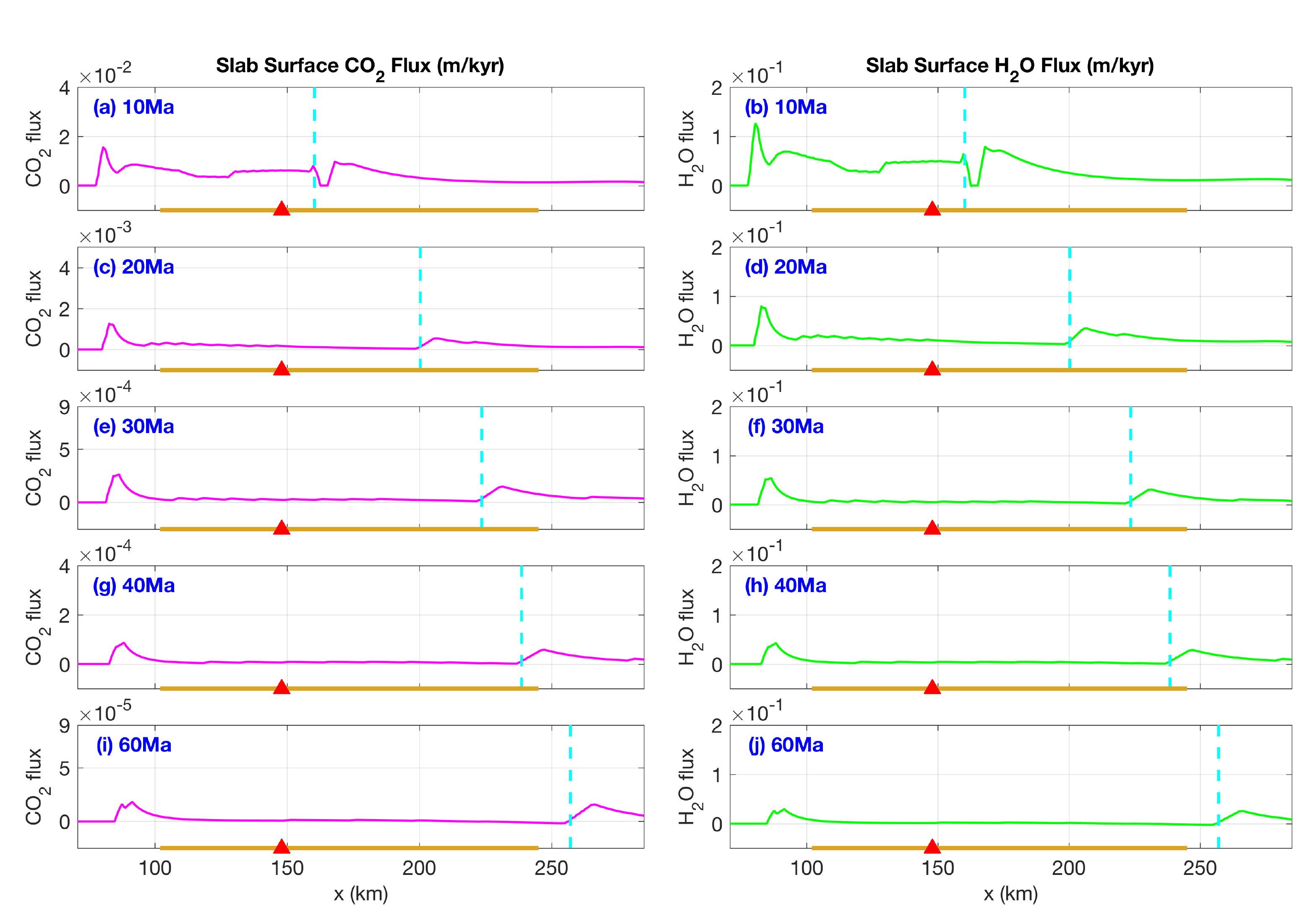}
\caption{CO$_2$ (left panels) and H$_2$O (right panels) fluxes along the surfaces of slabs with different ages. The result from the reference model with a 10-Ma-old slab is also plotted in the top row for comparison. Note the shift of the position for the onset of slab mantle devolatilization and the flux peak associated with it. Also note that the $y$-axis scales decrease from top row to bottom for CO$_2$ fluxes. }
\label{fig: age-surfacefluxcomp}
\end{figure}

Published closed-system and 1D open-system studies \citep{Kerrick:1998aa, Kerrick:2001aa, Kerrick:2001ab, Gorman:2006aa, Hacker:2008aa, Keken:2011aa} show that high-temperature subduction geotherms promote slab devolatilization and thus increase slab-surface volatile fluxes. We test this hypothesis using our 2D open-system model. Results with flow angle $\theta = 90^{\circ}$ and slab ages ranging from 10Ma to 60Ma are shown in Figure \ref{fig: age-surfacefluxcomp}, where it can be seen that, for old slabs with cold geotherms, devolatilization is deferred to greater depths. For the coldest slab (60Ma old) in this study, there is almost no dehydration or decarbonation in the range of subarc depths. In general, relative to the 10-Ma slab in the reference model, old, cold slabs will release less volatiles around subarc depths, therefore promoting $\mathrm{CO_2}$ \& $\mathrm{H_2O}$ recycling into deep mantle \citep{Rupke:2004aa}.

Flow angles in Figure \ref{fig: age-surfacefluxcomp} are all $90^{\circ}$ and, according to our assessment of the directional effect of flows, it can be inferred that up-slab flows ($\theta > 90^{\circ}$) will produce flux peaks. However, the positions of these peaks are controlled by the position of the onset of slab-mantle devolatlization. Given that the onset of slab-mantle devolatlization is also deferred to greater depths in old slabs (Fig. \ref{fig: age-surfacefluxcomp}), up-slab flows would not affect the subduction efficiency of H$_2$O and CO$_2$ (see section \ref{sec: efficiency}). In other words, cold slabs promote H$_2$O and CO$_2$ recycling even if within-slab flows are upward along the slab.

\subsection{Effect of Serpentinization} \label{sec: hydration}
\begin{figure}[h!]
\centering
\includegraphics[scale=0.45]{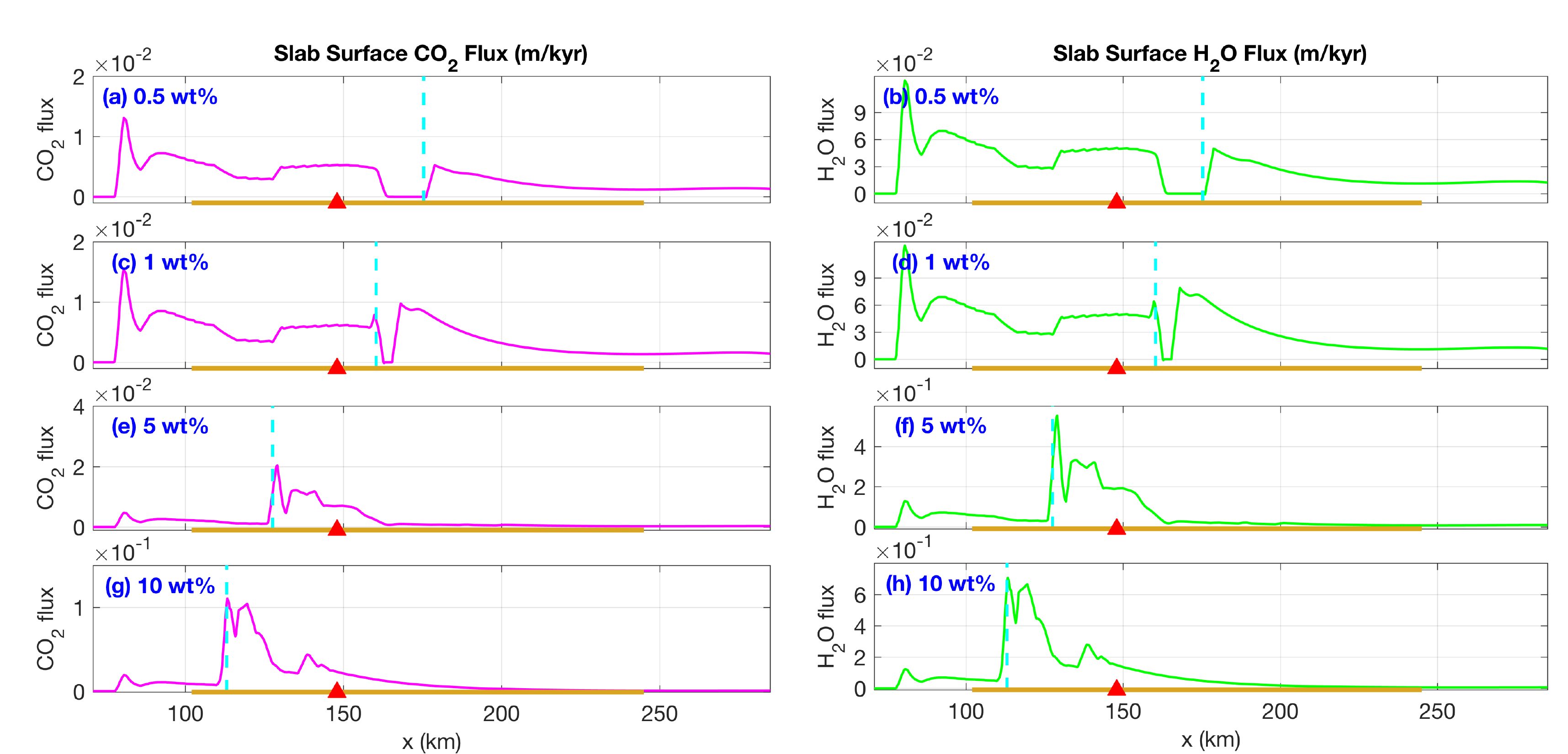}
\caption{Slab surface CO$_2$ (left panels) and H$_2$O (right panels) fluxes as a function of initial H$_2$O content in the basal layer. The reference model (section \ref{sec: refmodel}) contains 1 wt\% H$_2$O and its results are plotted in panels c \& d for comparison. The cyan dashed lines mark the onset of dehydration of this basal layer. Note the increase of $y$-axis scales in the lower panels. }
\label{fig: h2o-surfacefluxcomp}
\end{figure}

It is commonly proposed that hydration of slab lithospheric mantle takes place in fast-spreading oceans at the outer rise near the trench \citep{Ranero:2003aa}; subsequent dehydration leads to slab embrittlement, which has been invoked for interpreting subduction zone seismicity \citep{Kerrick:2002aa, Peacock:2001aa, Paulatto:2017aa, seno96}. Observational seismic studies infer the extent of serpentinization to be 5\%--31\% \citep{Garth:2014aa, Garth:2017aa, Grevemeyer:2007aa}, which corresponds to $\mathrm{H_2O}$ content of 0.6--3.5~wt\% for hydrated slab mantle. Given this uncertainty of $\mathrm{H_2O}$ content and its control on the within-slab flux pattern (e.g., Fig. \ref{fig: dir-flux}), we test how this factor affects slab volatile storage and surface fluxes by varying the $\mathrm{H_2O}$ content in the incoming slab mantle lithosphere from 0.5 wt\% ($\sim$4\% serpentinization) to 10 wt\% ($\sim$76\% serpentinization). Panels in Figure \ref{fig: h2o-surfacefluxcomp} show the resultant variation of $\mathrm{CO_2}$ and $\mathrm{H_2O}$ fluxes along the slab surface, where 1 wt\% H$_2$O is the value used for the reference model. Firstly, the position of onset for lithosphere devolatilization moves from depth of $\sim$125 km to $\sim$80 km as the hydration extent of incoming slab increases. This is because increasing the bulk H$_2$O content while keeping bulk CO$_2$ constant (negligibly small 0.02 wt\%) decreases the onset temperature of devolatilization. Secondly, the magnitudes of slab surface fluxes are substantially increased by the elevated basal $\mathrm{H_2O}$ content. Figure \ref{fig: h2o-surfacefluxcomp}e--h show that serpentinite dehydration can dominate other controls of slab surface $\mathrm{CO_2}$ and $\mathrm{H_2O}$ fluxes and thus can play a crucial role in arc magmatism and subduction-zone volatile recycling \citep{Shimoda:2019aa}. 

\begin{figure}[h!]
\centering
\includegraphics[scale=0.45]{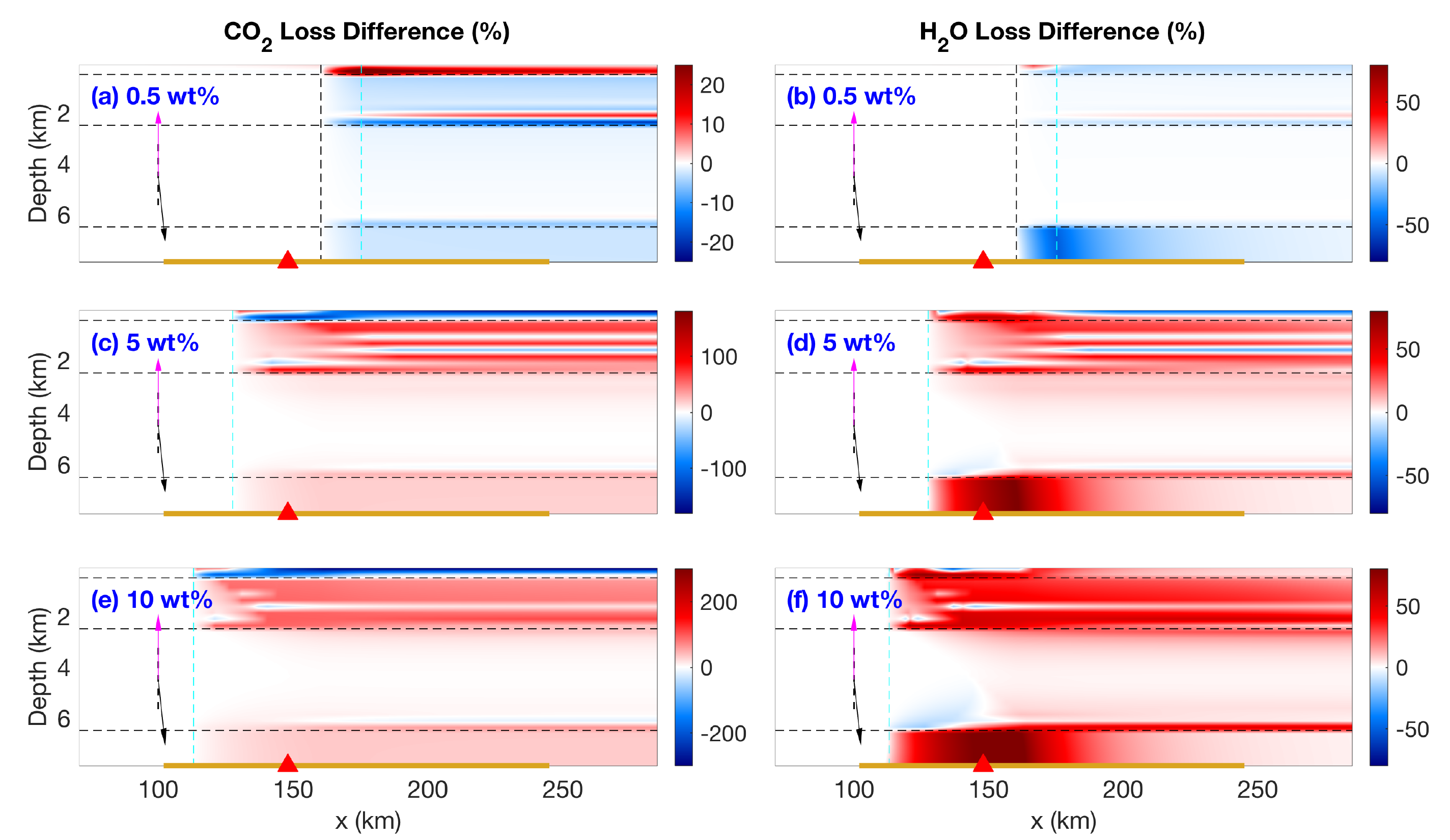}
\caption{Differences in CO$_2$ (left panels) and H$_2$O (left panels) loss relative to the reference model, caused by various initial $\mathrm{H_2O}$ content in the slab mantle. The percentage values are relative to the initial rock volatile content, as in Figure \ref{fig: diffopenclose}. Red color corresponds to enhanced volatile loss or diminished volatile gain, whereas blue color corresponds to diminished loss or enhanced gain. Note that the thin vertical black dashed lines in (a) and (b) mark the onset of dehydration in hydrated slab mantle for the reference model, and are plotted here for better comparison. }
\label{fig: h2o-volatileloss}
\end{figure}

In Figure \ref{fig: h2o-volatileloss}, the differences in CO$_2$ and H$_2$O loss relative to that of the reference model are mapped for different hydration states of the slab mantle. For the smaller extent of hydration (Fig. \ref{fig: h2o-volatileloss}a--b), the weakened infiltration leads to general reduction in reaction progress, as demonstrated by the decrease in both the CO$_2$ gain at the sites of carbonation (red stripes in Fig. \ref{fig: h2o-volatileloss}a), and the CO$_2$ loss at the sites of decarbonation (blue stripe/regions in Fig. \ref{fig: h2o-volatileloss}a). In particular, there is a stripe pair above the basalt--gabbro interface in Figure \ref{fig: h2o-volatileloss}a, but vertically in reverse order to that in Figure \ref{fig: diffopenclose}a. This is still attributable to the diminished infiltration flux as explained above. By contrast, for higher extent of slab mantle hydration (Fig. \ref{fig: h2o-volatileloss}c--f), stripe pairs similar to those in Figure \ref{fig: diffopenclose}a also appear in Figure \ref{fig: h2o-volatileloss}b \& e, but are pushed to upper levels due to the increased infiltration flux. Moreover, for very strong infiltration as in Figure \ref{fig: h2o-volatileloss}e, decarbonation is so strong that even the carbonation within the basalt layer is offset (white stripe at $\sim$2 km in Fig. \ref{fig: h2o-volatileloss}e). If inspected closely, a red--blue stripe pair appears at the peridotite-gabbro interface at $\sim$6.5 km in Figure \ref{fig: h2o-volatileloss}e, which is not the case for the reference model in Figure \ref{fig: diffopenclose}a, and the reason is simply that the stronger infiltration makes the stripe pair discernible.

Quantitatively, for the intensively infiltrated basalt layer in Figure \ref{fig: h2o-volatileloss}c \& e, $\mathrm{CO_2}$ loss can be enhanced to as high as 100\%, whereas for the even more intensively infiltrated sedimentary layer, the rise in $\mathrm{CO_2}$ gain can be as high as 300\%. In other words, enhanced infiltration extracts an extra amount of $\mathrm{CO_2}$ from the gabbro and basalt layers along flow pathways, but much of the extracted $\mathrm{CO_2}$ is merely redistributed into the top sedimentary layer. By comparing Figure \ref{fig: h2o-surfacefluxcomp}c and e, or \ref{fig: h2o-surfacefluxcomp}c and g, it can be inferred that not all of the infiltration-induced CO$_2$ transfer is deposited in the sedimentary layer, a fraction of it escapes the slab top and contributes to the surface flux increase.

$\mathrm{H_2O}$ loss or gain differs from that of $\mathrm{CO_2}$. For the basalt and gabbro layers, both are infiltrated by H$_2$O-rich fluids that cause decarbonation above the layer contacts. As illustrated in the companion paper Part I, fractionation caused by decarbonation leads to a larger extent of overall devolatilization. The enhanced H$_2$O loss at depths $\sim$2.5 km and $\sim$6.5 km in Figure \ref{fig: h2o-volatileloss}d \& f is a result of this. For the top sedimentary layer, however, it is infiltrated by flow down a CO$_2$ concentration gradient, so the fractionation caused by carbonation leads to inhibited H$_2$O loss, as reflected in Figure \ref{fig: h2o-volatileloss}d. Different from that in Figure \ref{fig: h2o-volatileloss}d, the top sedimentary layer in Figure \ref{fig: h2o-volatileloss}f experiences more H$_2$O loss. This is because the stronger infiltration down the CO$_2$ gradient also induces dehydration that counteracts the fractionation effect that is dominant for the case in Figure \ref{fig: h2o-volatileloss}d.

It is worth noting that field studies document the occurrence of a metasedimentary layer immediately on top of serpentinized slab mantle, where the sediments firstly experience decarbonation\slash hydration and secondly carbonation\slash partial dehydration \citep{Brovarone:2014aa, Piccoli:2016aa}. The flow in the outcrops are inferred to be confined into up-slab fluid channels. This field-documented two-stage flow event can be roughly elucidated by the current model. Taking Figure \ref{fig: h2o-volatileloss}e for illustration, if there are channels tapping fluids from dehydrating mantle lithosphere in an up-slab direction (e.g., Fig. \ref{fig: dir-flux}d), then the subducting sediments will be infiltrated by very H$_2$O-rich fluids, leading to the first stage decarbonation\slash hydration. At greater depths, when the tapped fluids scavenge CO$_2$ by passing through gabbro and basalt layers, carbonation reaction can take place in the sedimentary layer, accounting for the second-stage flow documented in the field. 

\subsection{Effect of Sediment Removal} \label{sec: diapirism}
For all of our models so far, a sedimentary layer of 500~m is assumed to cover the slab top and not detach during subduction. However, the sediments are subject to off-scraping in the accretionary prism of subduction zones \citep[e.g.,][]{Raymond:2019aa}. In addition, the slab surface sediments that survive off-scraping at trenches are further subject to removal by diapirism. Previous studies \citep{Behn:2011aa, Marschall:2012aa} showed that a sedimentary layer $>$100 m thick and $\sim$200 $\mathrm{kg \, m^{-3}}$ lighter than surrounding peridotites may form rising diapirs that traverse the mantle wedge. \citet{Kelemen:2015aa} propose that diapirs entraining carbonates are an important avenue for carbon release to mantle lithosphere. We therefore assess the impact of sediment removal on slab volatile storage and emission based on a simplified model setup: (i) the topmost 500-m-thick sedimentary layer in the reference model is replaced by basalt of the same composition as the underlying layer, mimicking the scenario of complete removal of sediments; (ii) this replacement basaltic layer is excluded during calculations of the resultant volatile fluxes and subduction efficiency (see next section). (iii) fluid flow angles are varied among 22.5$^{\circ}$, 45$^{\circ}$, 90$^{\circ}$, 135$^{\circ}$, 157.5$^{\circ}$ to examine the resultant slab surface fluxes.

\begin{figure}[h!]
\centering
\includegraphics[scale=0.45]{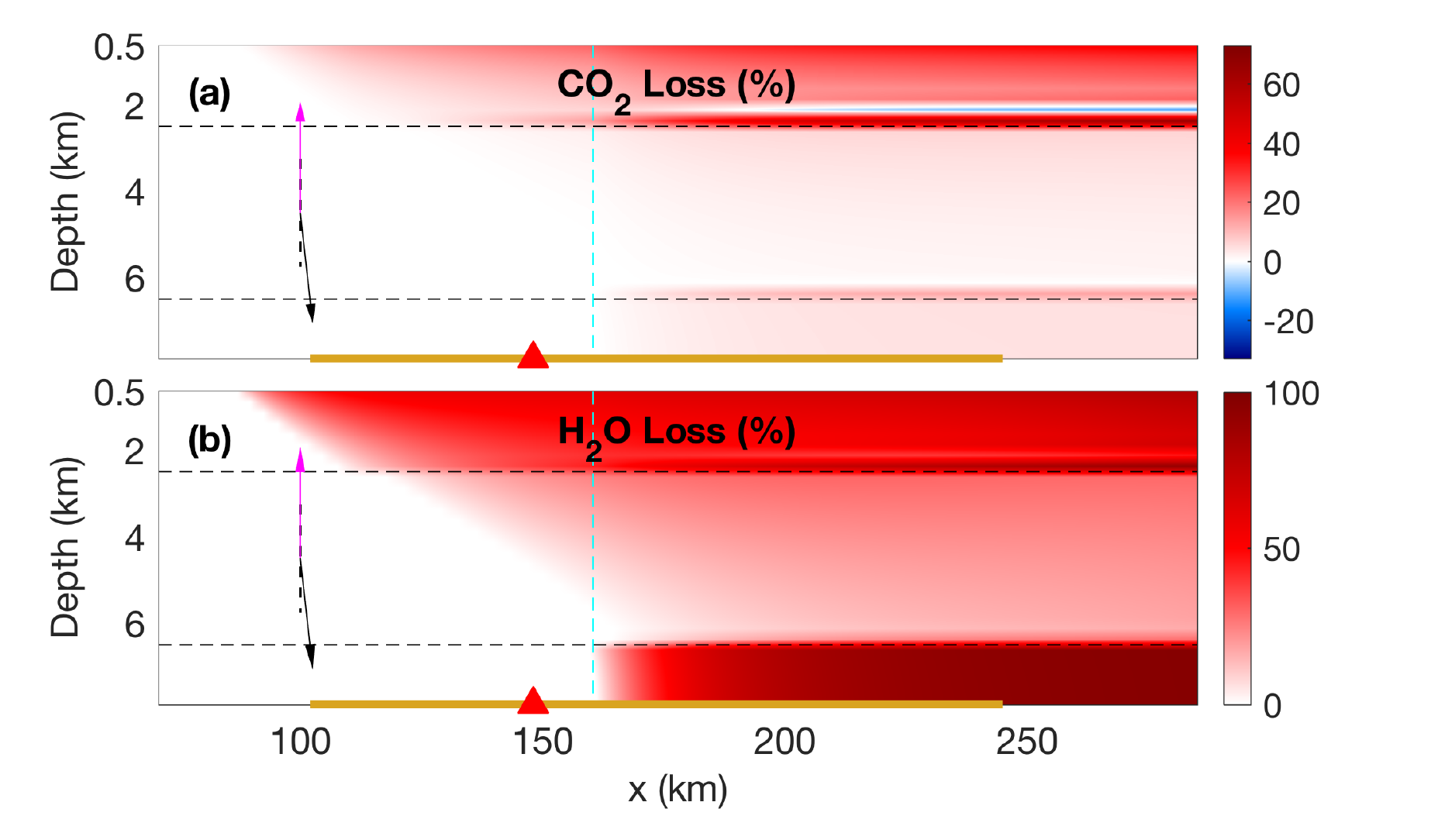}
\caption{Volatile loss for the three-layer slab model with flow angle $\theta=90 ^{\circ}$, that is, there is no capping sedimentary layer. Note that the depth starts from 0.5 km because the top layer is excluded from calculation. }
\label{fig: sed-volatileloss}
\end{figure}

For this three-layer model with a 90$^{\circ}$ flow angle, Figure \ref{fig: sed-volatileloss} maps the volatile loss within the slab. Regarding CO$_2$ loss (Fig. \ref{fig: sed-volatileloss}a), except for the thin sheet of CO$_2$ gain above the basalt--gabbro interface, the entire plate loses CO$_2$, and the loss can reach a maximum of $\sim$70\% at the infiltration site from gabbro to basalt. The thin sliver of up to 25\% CO$_2$ gain is due to the infiltration-related carbonation, as previously explained. In contrast, H$_2$O loss in Figure \ref{fig: sed-volatileloss}b is similar to that of the reference model. This is because the sedimentary layer, even if it survives the removal, should dehydrate and not restrict $\mathrm{H_2O}$ transport from beneath.

\begin{figure}[h!]
\centering
\includegraphics[scale=0.4]{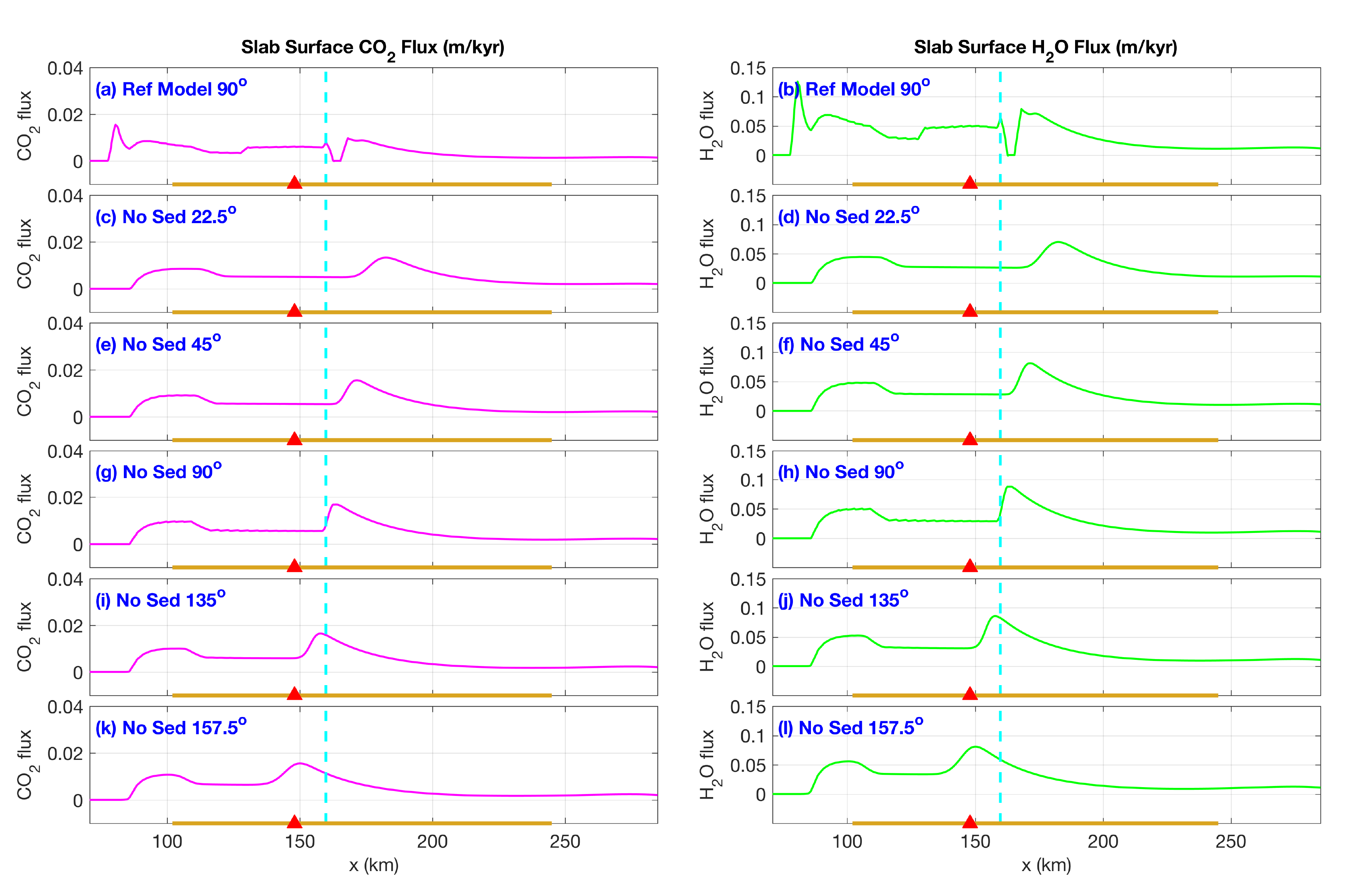}
\caption{$\mathrm{CO_2}$ (left panels) and $\mathrm{H_2O}$ (right panels) fluxes on the slab surface. Panels (a) and (b) are from the reference model that has a 500-m-thick sedimentary layer atop the slab, whereas the rest are results from the three-layer model with different flow angles ($\theta$).}
\label{fig: sed-surfaceflux}
\end{figure}

Figure \ref{fig: sed-surfaceflux} provides the slab-surface CO$_2$ and H$_2$O flux profiles for the three-layer model with the various flow angles ($\theta$). It shows that removal of the capping sediments leads to an overall increase of the CO$_2$ fluxes, and additionally to the emergence of a broad flux peak at the subarc depth range. In particular, when fluid flow is nearly up-slab (Fig. \ref{fig: sed-surfaceflux}k \& l), the position of the flux peak coincides with the average position of global arc volcanoes projected onto slab surface \citep{Syracuse:2006aa}. It is also interesting to compare between the flux profiles of the four-layer (Fig. \ref{fig: sed-surfaceflux}a) and three-layer models (Fig. \ref{fig: sed-surfaceflux}g), both with a flow angle of 90$^{\circ}$: the carbonation reaction in the sedimentary layer clearly suppresses the CO$_2$ flux peak that would appear without it. Up-slab flows ($\theta>90^{\circ}$) can therefore avoid the more reactive sediments at greater depths, and eventually contribute to slab surface $\mathrm{CO_2}$ fluxes. On the other hand, removal of the capping sediments does not increase the slab surface H$_2$O fluxes, except that it removes the flux dip at $\sim$160 km in the four-layer reference model (Fig. \ref{fig: sed-surfaceflux}b). It is clear from our analysis of the open-system reference model (section \ref{sec: refmodel}) that the flux dip stems from porosity reduction due to enhanced carbonation in the sedimentary layer, thus the dip is eliminated after the removal of the sedimentary layer. 

The simplified three-layer model represents an end-member case where the subducted slab is devoid of sediments. By comparing the total CO$_2$ and H$_2$O amount in the three-layered slab with those in the four-layered slab in the reference model, we find that sediment removal causes $\sim$11 \% CO$_2$ loss and $\sim$9 \% H$_2$O loss from the slab. It is noteworthy that these estimates of CO$_2$ and H$_2$O loss via sediment removal do not account for the thermal effect resulted from the direct contact between the basalt layer and mantle wedge. Further heating of the basalt layer by mantle wedge can increase the CO$_2$ and H$_2$O loss from the slab, and thus their fluxes along the slab surface.

\subsection{Subduction Efficiency of H$_2$O and CO$_2$} \label{sec: efficiency}
\begin{figure}[h!]
\centering
\includegraphics[scale=0.4]{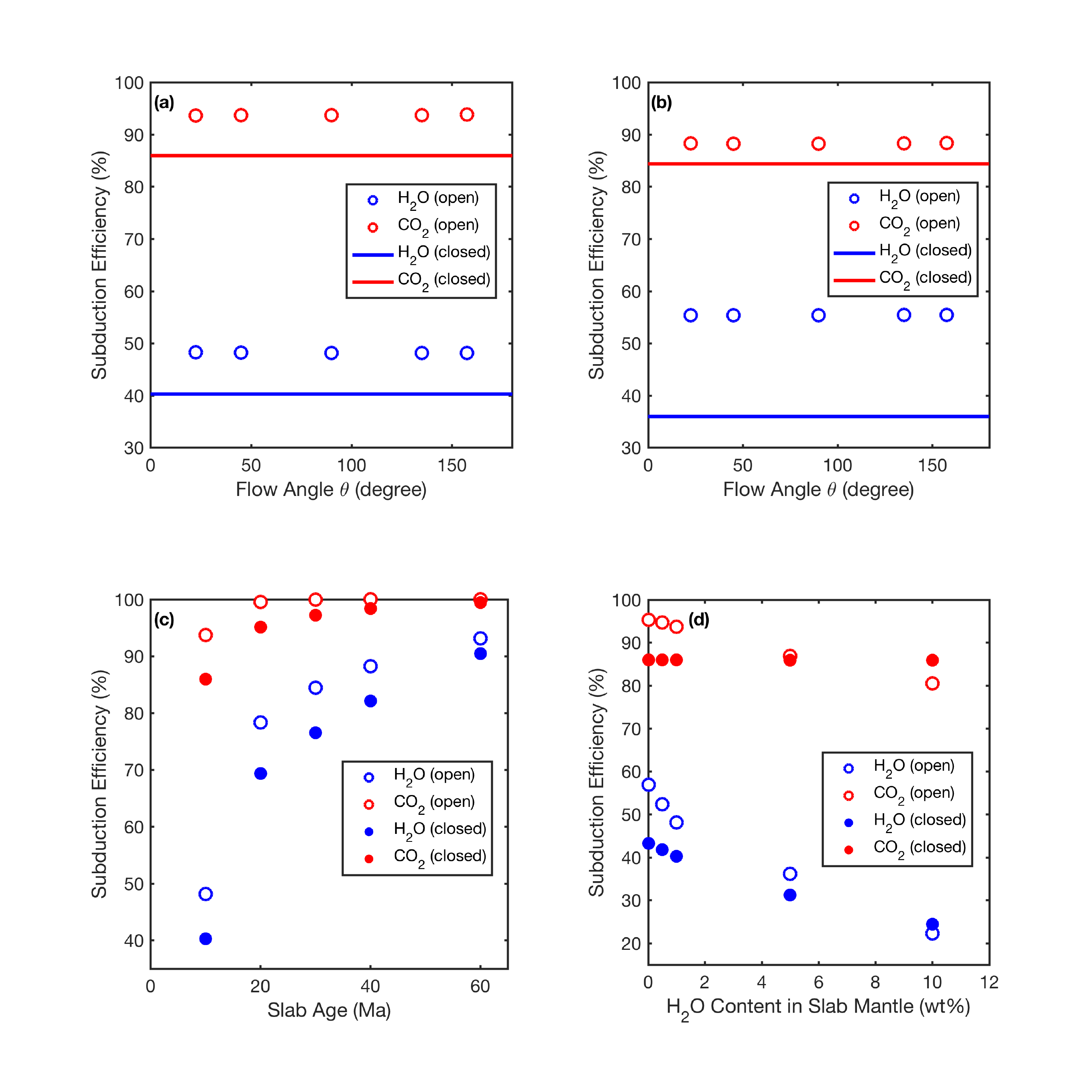}
\caption{Subduction efficiency of H$_2$O (blue) and CO$_2$ (red). (a) Effect of flow angle $\theta$, results extracted from section \ref{sec: refmodel} and \ref{sec: flowdirection}. The closed-system subduction efficiencies do not depend on flow directions, so are plotted as lines. (b) Effect of flow angle $\theta$, results extracted from the model without the top sedimentary layer in section \ref{sec: diapirism}. Same as in (a), the horizontal lines denote closed-system efficiencies. (c) Effect of slab age, results extracted from section \ref{sec: age}. Open circles denote open-system results, whereas dots denote closed-system ones; same markers in (d). (d) Effect of the H$_2$O content in the hydrated slab upper mantle, results extracted from section \ref{sec: hydration}. }
\label{fig: sub-efficiency}
\end{figure}

The subduction efficiency of H$_2$O (and CO$_2$) in this study is defined as the quotient of its total mass residing in the solid rock phase beyond $\sim$200 km deep over that residing in the incoming slab. To demonstrate how the various factors in the preceding sections affect the subduction efficiency, we plot them in Figure \ref{fig: sub-efficiency}. 

Volatile redistribution among different layers shown in Figure \ref{fig: h2o-volatileloss} implies that the subduction efficiency for individual layers depends on flow direction; however, Figure \ref{fig: sub-efficiency}a and b suggest that flow direction exerts almost no influence on the subduction efficiency for the entire slab. Comparison between the open-system results in Figure \ref{fig: sub-efficiency}a and b indicates that the subduction efficiency of CO$_2$ is reduced by $\sim$7\%, due to the absence of a sedimentary layer where carbonation reaction can occur. Figure \ref{fig: sub-efficiency}c illustrates that slab age strongly influences volatile subduction efficiency---the older the slab, the higher the efficiency. In particular, the 60-Ma-old cold slab is nearly 100\% and 93\% efficient in recycling CO$_2$ and H$_2$O beyond mantle depth of $\sim$200 km. In addition, Figure \ref{fig: sub-efficiency}d shows that the H$_2$O content of hydrated slab mantle exerts a much stronger control on the efficiency of volatile subduction. For extensively hydrated slab mantle (e.g., 10 wt\% H$_2$O or $\sim$76 \% serpentinization), the recycling efficiency of CO$_2$ can be reduced to $\sim$80\% and that of H$_2$O to as low as $\sim$21\%. 

Given that reactive flows are path-dependent, the finding that the subduction efficiency in Figure \ref{fig: sub-efficiency}a and b is insensitive to flow direction is surprising. However, it can be understood from the two assumptions in the model. The first is the local equilibrium assumption, which dictates that, as long as the $P$--$T$ structure of the slab is constant, steady-state volatile production within the entire slab is determined by $\rho_s \, v_s \partial \phi / \partial z$ (see Appendix \ref{apx:a}). Since porosity $\phi$ further depends on bulk composition, which is affected by flow directions, in theory the volatile production ($\rho_s \, v_s \partial \phi / \partial z$) should be path-dependent and thereby give rise to path-dependence of subduction efficiency. However, changes in bulk H$_2$O and CO$_2$ content induced by varying flow direction are small and spatially limited such that the path-dependence of subduction efficiency becomes indiscernible in Figure \ref{fig: sub-efficiency}a and b. Thus, the flow angle ($\theta$) shifts the volatile flux peaks, but not the overall amount of devolatilization. The second relevant model assumption is the uniform flow direction across slab, which makes strongly focused flows unlikely and hence prevents substantial alterations of bulk volatile content. 

The subduction efficiency of H$_2$O has also been estimated by previous studies that don't consider CO$_2$. For example, \citet{Keken:2011aa} estimate an average subduction efficiency of 32\% for H$_2$O in the present-day subducting slabs. In their model, the basal slab-mantle layer is 4 km thick and contains 2 wt\% H$_2$O. In terms of the total amount of H$_2$O infiltration, this is roughly equivalent to a H$_2$O content of 5.3 wt\% in the 1.5 km thick basal layer in our model. As shown in Figure \ref{fig: sub-efficiency}d, our model would yield a H$_2$O subduction efficiency of $\sim$35\%. Nonetheless, the average slab age in the model by \citet{Keken:2011aa} is older than 10 Ma that is adopted in the calculation for Figure \ref{fig: sub-efficiency}d, implying that our model would yield a H$_2$O subduction efficiency higher than 35\% for the present-day global subduction zones. The higher subduction efficiency of H$_2$O is expected because our model accounts for open-system effects. As demonstrated in Figure \ref{fig: sub-efficiency}, the subduction efficiency of H$_2$O is generally higher in open-system models than in closed-system ones. This is mainly due to the fractionation-induced inhibition of H$_2$O loss in the bulk interior of subducted lithological layers (see section \ref{sec: refmodel}). Only in the case of strong infiltration in Figure \ref{fig: sub-efficiency}d does the H$_2$O (and CO$_2$) subduction efficiency become lower in open systems than in closed systems. Similar to H$_2$O, CO$_2$ subduction efficiency is also elevated by open-system fractionation. Moreover, comparison between Figure \ref{fig: sub-efficiency}a and b suggests that, for closed systems, removal of top sedimentary layer exerts almost no effect on the CO$_2$ subduction efficiency; however, for open systems, the CO$_2$ subduction efficiency is reduced by $\sim$7\% due to the sediment removal, highlighting again the importance of open-system effects.

\section{Discussion} \label{sec: discussion}
By treating the subducting slab as a non-deformable plate and assuming local equilibrium and uniform flow direction, our model solves for the steady-state reactive flows of H$_2$O and CO$_2$ and quantifies their fluxes from and storage within the slab. The merit of the model is its capability to couple open-system thermodynamics with fluid flow in a more realistic and consistent way. However, in achieving this capability, the choices made in model setup also impose limitations.

The first limitation lies in neglecting partial melting. Various studies have focused on the effects of $\mathrm{H_2O}$ on partial melting, and suggest that melting primarily occurs in the overlying mantle wedge due to solidus depression caused by slab-derived $\mathrm{H_2O}$ \citep{Grove:2006aa, Grove:2012aa}. Field studies in volcanic arc settings, however, discovered adakites that are interpreted to be evidence of slab melting in hot geotherms \citep{Sen:1994aa, Drummond:1996aa}. When $\mathrm{CO_2}$ is taken into account, experimental studies suggest that even in hot subduction zones, modern slab geotherms are still below the solidi of carbonated basalts and peridotites, and that only carbonated sediments are promising for carbonatite liberation \citep[and references therein]{Dasgupta:2013aa}. On the other hand, \citet{Poli:2015aa} shows that partial melting of hydrous carbonated gabbros can produce carbonatitic liquids in warm subduction zones, as long as the gabbros are Ca-rich, a result likely from extensive hydrothermal alteration. In light of these experimental findings, neglecting the partial melting of slab-mantle peridotite in our model is a reasonable choice, but ignoring the melting of mafic and sedimentary layers could overestimate the subduction efficiency of CO$_2$, especially when migration of carbonatitic melt can remove considerable CO$_2$ from the slab. Moreover, absence of partial melts in the current model indicates that the thermal feedbacks from associated latent heat and melt transport is unaccounted for \citep[e.g.,][]{Jones:2018aa}. The influences of perturbation to the slab thermal structure caused by the loss of partial melts and\slash or sediments (section \ref{sec: diapirism}) need to be further assessed in future research.

The second limitation stems from the binary H$_2$O--CO$_2$ molecular fluids assumed for the liquid phase in our model. Variations of pressure, temperature and particularly oxygen fugacity during subduction can give rise to the dominance of methane (CH$_4$) or precipitation of graphite \citep[e.g.,][]{Galvez:2013aa,Brovarone:2017aa}. Moreover, ionic carbon species (e.g., HCO$_3^{-}$, CO$_3^{2-}$) are inferred to be common in subduction zone settings \citep[e.g.][]{Frezzotti:2011aa, Ague:2014aa}, although detailed thermodynamic modelling suggests that molecular carbon species can dominate over ionic ones in fluids derived from subducted basalts \citep{Galvez:2016aa} and sediments \citep{Connolly18}. The latter study also shows that potassium in ionic form can be depleted from the slab-top sediments, destablizing hydrous phases like micas and thus facilitating volatile release. In addition to potassium, the activity of dissolved silica and aluminum also affects the stability of silicates that participate in carbonate dissolution reactions \citep{Galvez:2019aa}. Therefore, omission of the direct contribution from carbon species other than CO$_2$ and indirect contribution from dissolved non-volatile elements (e.g., Si and Al) might underestimate the water and carbon fluxes from the slab, and in turn overestimate their recycling efficiency. 

The third limitation arises from the assumption of uniform flow directions. As explained in section \ref{sec: setup} on the model setup, choosing the flow direction $\theta$ as a free parameter enables the model to explore the variation of general patterns of within-slab flows predicted by various dynamic models; however, it suffers from losing the details of flow dynamics within the slab. In particular, the strong flow focusing, widely documented in the field \citep[e.g.,][]{Ague:2007aa, Breeding:2003aa, Philippot:1991aa, Barnicoat:1995aa, Piccoli:2016aa, Piccoli:2018aa}, and predicted by dynamic models \citep[e.g.,][]{Faccenda:2009aa, Malvoisin:2015aa, Plumper:2017aa, Wilson:2014aa}, cannot be simulated in the current model. As demonstrated by \citet{Wada:2012aa}, heterogeneous hydration associated with local flow focusing can promote H$_2$O release and reduce its subduction efficiency. Therefore, further work remains to incorporate dynamics into the open-system flow model to evaluate how flow focusing can affect the slab-surface fluxes of CO$_2$ and H$_2$O, and their recycling efficiency by subduction.

The fourth limitation relates to the assumption of no compaction in the slab. To assess the effect of compaction, we compare the magnitude of fluid fluxes in our reference model with that in the compaction model by \citet{Wilson:2014aa}. Our slab model (section \ref{sec: refmodel}) initially contains $\sim$2.60 wt\% H$_2$O on average, yielding a maximum surface flux of $\sim$0.1 m kyr$^{-1}$ (Fig. \ref{fig: surfacefluxcomp}f). In contrast, the compaction model initially contains $\sim$1.85 wt\% H$_2$O on average, yielding a maximum surface flux of $\sim$0.8 m kyr$^{-1}$. Taken at face value, the compaction model generates a flux that is eight times larger than that from our non-compaction model. However, as shown in Figure \ref{fig: diffopenclose}b, the H$_2$O flux drop in our reference model is caused not only by neglecting compaction, but also by the open-system fractionation, which can account for up to $\sim$70\% of the flux drop. If this open-system effect were considered by \citet{Wilson:2014aa}, the magnitude of H$_2$O fluxes in their model would become comparable to that in our reference model, indicating that the influence of compaction is insignificant with respect to the magnitude of slab surface fluxes. Nevertheless, as stressed above, compaction may be important if it is a control on fluid flow pathways within the slab. 

In interpreting our model results on the slab surface fluxes, an additional assumption is made that magmas produced by flux melting of the mantle wedge traverse the wedge vertically without any lateral deflection to feed arc volcanoes. It follows that, if the average position of global arc volcanoes is projected downwards onto the slab surface, it should correspond to a peak in slab surface flux. It is under this assumption that the up-slab flows in Figure \ref{fig: surfacefluxcomp} are necessary to produce the melt feeding arc volcanism. However, two-phase geodynamic models of the mantle wedge suggest that the melt and fluid transport through the wedge is more complex than simply vertical, and depends on many factors including solid rheology and grain size in the wedge \citep[][]{Wilson:2014aa, Cerpa:2017aa, Cerpa:2018aa}. In addition, the coincidence of flux peaks and the projected volcanic arc position in Figure \ref{fig: surfacefluxcomp} and \ref{fig: sed-surfaceflux} is premised on the hot thermal structure of the 10-Ma-old slab. As demonstrated in Figure \ref{fig: age-surfacefluxcomp} and \ref{fig: h2o-surfacefluxcomp}, old slab shifts the flux peak to greater depths, and higher hydration extent of the slab base shifts the peak to shallower depths. Since slower convergence rate leads to a warmer subducting slab \citep{Penniston-Dorland:2015aa}, it can be expected that the position of the flux peak also varies with convergence rate. How the shift in flux-peak position influences the dynamics of melt or volatile transport in the wedge, and eventually the genesis of arc magmas, remains to be explored by future models that include both the dynamics and thermodynamics in the wedge. 

\section{Conclusion}
In summary, we assess the controlling factors for $\mathrm{CO_2}$ and $\mathrm{H_2O}$ release from subducting slabs. We find that up-slab flows are necessary to form a flux peak at subarc depths, which could be responsible for the magmatic genesis of arc volcanoes if the flux-induced melts traverse the mantle wedge vertically. Such a result aligns with the findings of dynamic studies by \citet{Wilson:2014aa} where a flow channel is formed to induce up-slab flows. Sufficiently hydrated slab mantle can relax the above requirement for flow directions, but when the slab mantle is additionally carbonated, the onset position of its devolatilization becomes deeper, still making up-slab flows important for supplying maximum volatile release at subarc depths. Moreover, H$_2$O infiltration from dehydrating slab mantle can mobilize CO$_2$ in the overlying gabbroic and basaltic layers, but only redistribute the carbon to the slab surface sediments where significant carbonation takes place. The subducting sediments, if passing the accretionary prism and surviving diapiric removal and partial melting, can be an important barrier to CO$_2$ liberation from slabs. Conversely, loss or partial melting of the sediments can be crucial avenues for CO$_2$ transfer from subducting slab to mantle wedge. The subduction efficiency of H$_2$O is $\sim$20\%--90\% in our model, generally higher than that in previous closed-system models that consider only H$_2$O. The CO$_2$ subduction efficiency is $\sim$80\%--100\%. The higher volatile subduction efficiency compared to closed-system models stems from the open-system fractionation effect that inhibits overall devolatilization in the interior of lithological layers. The absolute values are likely overestimates because the effects of flow focusing and fluid species other than H$_2$O and CO$_2$ are ignored in the model, which should be considered in future studies. 

%\appendix
%\section{}
\appendix
\section{Appendix} \label{apx:a}
\begin{figure}[h!]
\centering
\includegraphics[scale=0.8]{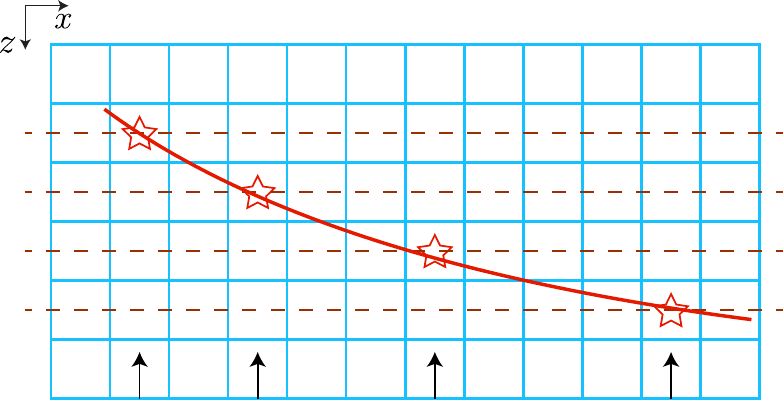}
\caption{An illustrative part of the numerical grid in this study. Stars mark the position where onset of devolatilization occurs at relevant slab depths, and the connecting red solid line is the envelope for devolatilization onset within the slab. Brown dashed lines are iso-depth traverses along which $\partial \phi/ \partial x$ will be plotted in Figure \ref{fig: profiles}a, and black arrows denote the direction of flux integration for the reference model ($\theta$=90$^{\circ}$). }
\label{fig: grid}
\end{figure}

At steady state, the fluid fluxes can be calculated according to equation \eqref{eq: bulkmass} that further reduces to:
\begin{equation} \label{eq: fluxeq}
\rho_f \nabla \cdot ( \boldsymbol{v}_f\phi )= \rho_s \boldsymbol{v}_s \cdot \nabla \phi.
\end{equation}
Figure \ref{fig: grid} shows part of the numerical grid established on the slab coordinate system (Fig. \ref{fig: geometry}b), where the fluid fluxes $F=v_f \phi$ can then be expressed through integration along flow path ($l$): 
\begin{equation} \label{eq: fluxint}
F = \int_{\mathrm{path \ start}}^{\mathrm{path \ end}} \frac{\rho_s}{\rho_f} v_s \frac{\partial \phi}{\partial x} \ \mathrm{d} l.
\end{equation}

\begin{figure}[h!]
\centering
\includegraphics[scale=0.7]{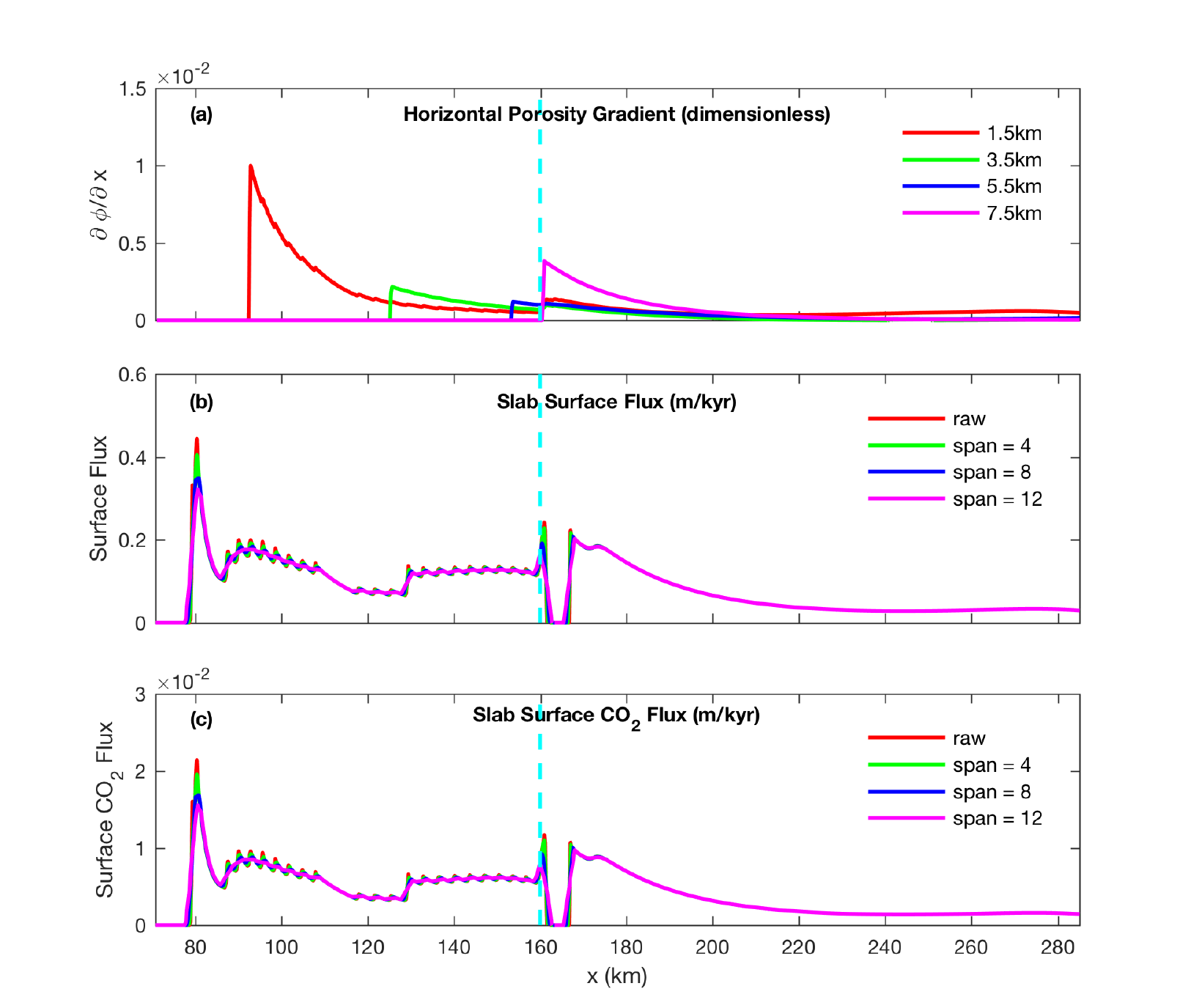}
\caption{(a). Profiles of $\partial \phi/ \partial x$ that contributes to fluid fluxes. Different line colors are used to distinguish the depths at which these profiles are extracted from the open-system reference model in section \ref{sec: refmodel}. (b). Slab surface fluxes from the reference model. Red line denotes the raw model result that shows flux fluctuations, whereas other lines represent the model result smoothed to different extent. The ``span" here means the number of neighboring grid points used for averaging the flux at a specific grid point. The spacing between grid points in the model is 250 m, so a span of 12 corresponds to an averaging window width of 3 km. (c). Same as in (b), but is for slab surface $\mathrm{CO_2}$ flux. Vertical dashed lines mark the onset of dehydration for slab lithospheric mantle. }
\label{fig: profiles}
\end{figure}

For the reference model with flow direction $\theta = 90^{\circ}$, flow paths are perpendicular to the slab, so the integration is along the $z$-axis direction in Figure \ref{fig: grid}. At any specific depth (dashed lines in Fig. \ref{fig: grid}), the onset of devolatilization leads to an abrupt elevation of porosity ($\phi$) and thus a pulse in the $\partial \phi/ \partial x$ profile along the iso-depth traverse. Figure \ref{fig: profiles} shows the $\partial \phi/ \partial x$ and flux profiles along various iso-depth traverses within the modelled slab, and Figure \ref{fig: profiles}a illustrates that the pulses in $\partial \phi/ \partial x$ appear at different horizontal positions in the slab. According to equation \eqref{eq: fluxeq}, when flow trajectories pass the locations of the pulses, the calculated fluid fluxes will inherit these pulses. For the reference model where flow paths are normal to $x$-axis and the pulses are horizontally apart from one another (Fig. \ref{fig: grid}), the inherited pulses lead to the fluctuation in the fluid flux distribution (Fig. \ref{fig: profiles}b--c and Fig. \ref{fig: refmodel}g). Beyond the slab distance $\sim$160 km where the basal slab mantle dehydrates, the envelope of the onset of devolatilization (red solid line in Fig. \ref{fig: grid}) is surpassed, so there are no pulses in $\partial \phi/ \partial x$ at any depth. The flux fluctuations disappear accordingly as in Figure \ref{fig: profiles}b--c and Figure \ref{fig: refmodel}h. 

The analysis above indicates that refining numerical grid will produce more but weaker fluctuations in slab surface fluxes before $\sim$160 km, but the trend and main features of flux profiles stay unchanged. To focus on the general trend and main features, we smooth the flux profiles by averaging over a span of neighboring grid points. As illustrated in Figure \ref{fig: profiles}b--c, smoothing with a span of 12 grid points (3 km) well preserves the flux trend and features, so it is applied to all the slab surface flux profiles in this study.

\clearpage

%\begin{acknowledgments}
%\end{acknowledgments}
\begin{acknowledgments}
We thank the three anonymous reviewers for their comments that helped us improve this manuscript. We thank the Isaac Newton Institute for Mathematical Sciences for holding the Melt in the Mantle program sponsored by EPSRC Grant Number EP/K032208/1. Support from Deep Carbon Observatory funded by the Sloan Foundation is acknowledged. M.T. acknowledges the Royal Society Newton International Fellowship (NF150745). D.R.J acknowledges research funding through the NERC Consortium grant NE/M000427/1, NERC Standard grant NE/I026995/1, and the Leverhulme Trust. This project has also received funding from the European Research Council (ERC) under the European Union's Horizon 2020 research and innovation programme (grant agreement n$^{\circ}$ 772255). This contribution is about numerical modelling, so it does not depend on experimental or field data. Relevant data and equations for reproducing the model results are already contained in the text. Nonetheless, we provide an example code on the usage of the thermodynamic parameterization: \url{https://bitbucket.org/meng_tian/example_code_thermo_module/src/master/}.
\end{acknowledgments}

\bibliography{manuscriptref}

\end{document}